\documentclass[10pt]{iopart}
\usepackage{setstack}\usepackage{iopams}
\usepackage{amssymb,graphicx}
\usepackage{citesort}

\usepackage[usenames]{color}

\begin{document}
\title{Critical Decay at Higher-Order Glass-Transition Singularities}
\author{W. G\"otze and M. Sperl}
\address{Physik-Department, Technische Universit\"at M\"unchen, 85747
  Garching, Germany} 

\begin{abstract}
Within the mode-coupling theory for the evolution of structural relaxation
in glass-forming systems, it is shown that the correlation functions for
density fluctuations for states at $A_3$- and $A_4$-glass-transition
singularities can be presented as an asymptotic series in increasing
inverse powers of the logarithm of the time $t$: $\phi(t)-f\propto \sum_i
g_i(x)$, where $g_n(x)=p_n(\ln x)/x^n$ with $p_n$ denoting some polynomial
and $x=\ln (t/t_0)$. The results are demonstrated for schematic models
describing the system by solely one or two correlators and also for a
colloid model with a square-well-interaction potential.
\end{abstract}

\submitto{\JPCM}
\today


\section{\label{sec:intro}Introduction}

Upon compression or cooling glass-forming liquids, there evolves a peculiar 
relaxation scenario called glassy dynamics. It is characterized by 
control-parameter sensitive correlation functions or spectra which are
stretched over large intervals of time $t$ or frequency $\omega$, 
respectively. The so-called mode-coupling theory (MCT) of ideal glass 
transition has been proposed \cite{Bengtzelius1984} as a mathematical model for
glassy dynamics. The basic version of that theory describes the system by $M$ 
correlation functions $\phi_q(t),\,q=1,\,2,\,\dots,\,M$, which 
have the meaning of canonically defined auto-correlators of density
fluctuations for wave vector moduli $q$ chosen from a grid of $M$ values. The 
theory deals with a closed set of coupled nonlinear equations of motion for 
the $\phi_q(t)$. The coupling coefficients in these equations are determined
by the equilibrium structure factors, which are assumed to be known smooth
functions of control parameters like temperature $T$ or density $\rho$. The
solution of the MCT equations describe a transition from an ergodic liquid 
state to a non-ergodic
glass state if the control parameters pass critical values $T_c$ or $\rho_c$,
respectively. This transition is accompanied with the appearance of a
dynamical scenario, whose qualitative features can be understood by asymptotic
solution of the equations for long times and control parameters close to
the critical values. The asymptotic formulas establish the universal features
of the glassy dynamics described by MCT. On the basis of this understanding,
one can construct schematic models. These are based on equations of
motion which have the same general form as those derived within the
microscopic theory of liquids, but use the number $M$ of correlators and the
coupling coefficients as model parameters. Thereby, one gets simplified models
whose results can be used for the analysis of data \cite{Goetze1991b}.

The ideal liquid-glass transition described by MCT is a fold bifurcation
exhibited by the equations of motion. One can show, that all singularities
that are generically possible, are of the cuspoid type
\cite{Goetze1995b,Franosch2002}. Using Arnol'd's notation \cite{Arnold1986},
an $A_l$ is a bifurcation which is equivalent to that for the roots of a real
polynomial of degree $l$. Simple schematic models using only a single
correlator exhibit besides the fold singularity also the cusp singularity
$A_3$ and the swallowtail singularity $A_4$ \cite{Goetze1991b}. The bifurcation
dynamics near a higher-order glass-transition singularity $A_l,\,l\geqslant 3$,
is
utterly different from the one near the liquid-glass-transition singularity of
type $A_2$. A major new feature is the appearance of logarithmic decay
yielding to a much stronger stretching than known for the $A_2$-scenario
\cite{Goetze1988b}. The result for the correlators of $M=1$ models has
been worked out in a certain leading-order multiple-scaling-law limit
\cite{Goetze1989d}. These formulas have been used to fit dielectric-loss data
of glassy polymer melts
\cite{Sjoegren1991,Flach1992,Halalay1996,Eliasson2001}, thereby providing some
hint that the MCT for higher-order singularities might be of
relevance for understanding glassy dynamics.

Let us consider a system of spherical particles interacting via a steep
repulsion potential characterized by a diameter parameter $d$, which is
complemented by an attractive potential. The latter shall be characterized by
an extension length $\Delta$ and an attraction-potential depth $u_0$. Such
system is specified conveniently by three control parameters: the packing
fraction $\varphi=\rho\pi d^3/6$, the dimensionless attraction strength
$\Gamma=u_0/(k_{\rm B}T)$ or the dimensionless effective temperature
$\theta=1/\Gamma$, and the relative attraction width $\delta=\Delta/d$. If
$\delta$ is sufficiently large, this potential is a caricature of a
van-der-Waals interaction. One gets a decreasing $\Gamma^c$-versus-$\varphi^c$
line of liquid-glass transitions in the $\Gamma$-$\varphi$ plane of the
thermodynamic states similar to what was first calculated within MCT for
Lennard-Jones systems \cite{Bengtzelius1986b}. 
For $\Gamma=0$, the mentioned line terminates at the critical
packing fraction $\varphi^c_{\rm HSS}$ for the vitrification of the
hard-sphere system. The decrease of $\varphi^c$ with increasing 
$\Gamma^c$ expresses the intuitive fact that cooling stabilizes the glass
state. If $\delta$ is sufficiently small, however, there appear two new
phenomena. First, for small $\Gamma$, the $\Gamma^c$-versus-$\varphi^c$ line
increases. Cooling stabilizes the liquid because bond formation creates
inhomogeneities which favor fluidity. The new liquid state for
$\varphi>\varphi^c_{\rm HSS}$ exhibits a reentry phenomenon. Glassification
occurs not only by increasing $\Gamma$, i.e. by cooling, but also by
decreasing $\Gamma$, i.e. by heating. Second, the
$\Gamma^c$-versus-$\varphi^c$ line can consist of two branches that form a
corner. The low-$\Gamma$-branch is terminated by the the
high-$\Gamma$-branch. The latter continues into the glass state as
glass-glass-transition line, which has an $A_3$-singularity as endpoint. These
two phenomena have been found by using Baxter's model for the structure
factor as input of the MCT equations \cite{Fabbian1999,Bergenholtz1999}. In
these calculations, the wave-vector cutoff $q_{\rm max}$ used in the MCT model
defines the range parameter $\delta=\pi/(q_{\rm max}d)$. The generic
possibility for a transition from small-$\delta$ states with
$A_3$-singularity to large-$\delta$ states without $A_3$-singularity is the
appearance of an $A_4$-singularity for some critical value $\delta^*$. Since
the $A_l$ bifurcations deal with topological singularities, the indicated
scenarios are robust, i.e., they occur for all potentials of the kind
specified above. The $A_4$-singularity was identified first for the
square-well system (SWS), i.e., for a system where a hard-core repulsion is
complemented by a shell of constant attraction strength $u_0$. Here,
$\delta^*\approx0.04$ was calculated \cite{Dawson2001}. The neighborhood of the
$A_4$ was analyzed for some other potentials with the conclusion that there
are no qualitative differences between results referring to different shapes
of the potential or to different approximation schemes for the structure
factor \cite{Goetze2003b}.

The above described systems with short-ranged attraction can be prepared as
colloidal suspensions. The liquid-glass-transition lines can be identified by
analyzing the nucleation processes. Light-scattering experiments can provide 
the density-correlation functions $\phi_q(t)$. Such studies
have identified the existence of liquid states for $\varphi>\varphi^c_{\rm
  HSS}$ and the reentry phenomenon \cite{Eckert2002,Pham2002,Pham2004}. 
Molecular-dynamics simulation studies can determine the mean-squared
displacement and the diffusivities with good accuracy. These quantities
exhibit drastic precursors of the liquid-glass transition. Several simulation
studies \cite{Pham2002,Foffi2002b,Zaccarelli2002b,Puertas2003} have confirmed 
the predictions on the reentry phenomenon. Near the corner formed by
large-$\Gamma$ and the small-$\Gamma$ transition lines, there should occur an 
almost
logarithmic decay of the density correlations $\phi_q(t)$, which is followed
by a von-Schweidler-law decay as beginning of an $\alpha$-relaxation process
\cite{Fabbian1999,Dawson2001}. Such scenario was first reported for micellar
solutions \cite{Mallamace2000}. This signature of the dynamics for
$\varphi>\varphi^c_{\rm HSS}$ states was detected also for colloidal
suspensions with depletion attraction \cite{Pham2004}.

In order to identify a higher-order singularity in data from experiment or 
from molecular-dynamics simulation, one has to identify the features of the 
correlators
$\phi_q(t)$ which are characteristic for these singularities. The general
theory of the logarithmic decay laws caused by an $A_l$ for $l\geqslant 3$, has
been developed and the relevant general scenarios have been illustrated for
schematic models \cite{Goetze2002,Sperl2004b}. The specific implications of
the general theory for the SWS, in particular the change of the features with
changes of the wave number and the peculiarities expected for the mean-squared
displacement, have been worked out as well \cite{Sperl2003a,Sperl2004}.
Simulation data for the tagged-particle-density correlators as function of the 
wave vector $q$ \cite{Puertas2002} provide a first hint that the predicted 
logarithmic decay processes for
$\varphi>\varphi^c_{\rm HSS}$ states near an $A_3$-singularity are present. 
Major progress was reported recently for simulation studies for two states of a
binary SWS \cite{Sciortino2003}. The logarithmic decay and its expected
deformation with wave-number changes has been detected convincingly. The 
identified amplitudes agree semi-quantitatively with the calculated ones
\cite{Sperl2003a}. In addition, the mean-squared displacement exhibits the 
expected control-parameter dependent power-law behavior. These findings provide
very strong arguments for the existence of a higher-order glass-transition 
singularity. One concludes that the cited MCT results on simple systems with 
short-ranged attraction reproduce some subtle features of glassy dynamics so
that further studies of these systems within that theory seem worthwhile.

If one shifts control parameters towards the ones specifying a higher-order
singularity, the time interval for logarithmic decay
expands. But, simultaneously, also the beginning of the time interval shifts to
larger values. Consequently, there opens a time interval of increasing length
between the end of the transient and the beginning of the logarithmic
decay. Within this interval, the correlators are close to the critical ones
$\phi^c_q(t)$, i.e., to the correlators calculated for control parameters at
the singularity. It should be expected that these critical correlators will be
detected in future data from experiments and from simulation studies. It was
shown for one-component schematic models that the critical correlators
approach their long-time limit proportional to $1/\ln^m(t/t_0)$, where 
$m=2/(l-2)$ for an $A_l$ \cite{Goetze1989d}. In the following, these results 
shall be extended in two directions. First,
the critical correlators shall be expanded in an asymptotic series so that an
estimate of the range of validity of various asymptotic formulas is
possible. Second, the $\phi_q^c(t)$ shall be calculated for the general theory
so that  a discussion of the $q$-dependent corrections of the leading
asymptotic formulas is possible for an $A_3$- and an $A_4$-singularity.

The paper is organized as follows. In Sec.~\ref{sec:generaleq}, the general
starting equations for an asymptotic discussion of critical relaxations are
compiled. Then, in Sec.~\ref{sec:crit_one_A3} and Sec.~\ref{sec:crit_one_A4},
the asymptotic expansion is carried out for one-component models for states
near an $A_3$- and $A_4$-singularity, respectively. The results will be
demonstrated quantitatively for schematic models. Section~\ref{sec:critA3}
shows how the theory for $\phi_q^c(t)$ for an $A_3$ can be reduced to the
theory of one-component models, The results are demonstrated for a
two-component schematic model and for the SWS. The analogous results for an
$A_4$-singularity are presented in Sec.~\ref{sec:critA4}. A summary is
formulated in Sec.\ref{sec:summary}.

\section{\label{sec:generaleq}General equations}

\subsection{\label{sec:A3A4}Equations for structural relaxation at 
glass-transition singularities}

Within the basic version of the mode-coupling theory for the evolution of
glassy relaxation (MCT), the system's dynamics is described by $M$ correlators
$\phi_q(t),\,q=1,\dots,M$. 
The theory uses the exact Zwanzig-Mori equations of motion. These
are specified by $M$ characteristic frequencies $\Omega_q>0$ and $M$
fluctuating-force kernels $M_q(t)$. The latter are decomposed into regular
terms $M^{\rm reg}_q(t)$ describing normal-liquid effects and in
mode-coupling kernels $m_q(t)$. The essential step in the derivation of
the theory is the application of Kawasaki's factorization approximation to
express the kernels $m_q(t)$ as absolutely monotone functions ${\cal F}_q$ of
the correlators. These functions depend smoothly on a vector $\mathbf{V}$ of
control parameters like density and temperature,

\begin{equation}\label{eq:def_m}
m_q(t) = {\cal F}_q [\mathbf{V},\phi_k(t)]\,.
\end{equation}
Vector $\mathbf{V}$ specifies the equilibrium structure functions of the
system. Using Laplace transforms of functions of time, say
$F(t)$, to functions defined in the upper plane of complex frequencies
$z$, $F(z) = i\int_0^\infty\,dt\,\exp(izt)\,F(t)$, the equations of motion
read $\phi_q(z) = -1/\{z-\Omega_q^2[z+M^{\rm reg}_q(z)+\Omega_q^2
m_q(z)]\}$ \cite{Goetze1991b}. Glassy dynamics is characterized by
long-time decay processes that lead to large small-frequency contributions
to $m_q(z)$. These small-$z$ contributions to $m_q(z)$ dominate over
$z+M^{\rm reg}_q(z)$. Therefore, glassy dynamics is described by the
simplified equation $\phi_q(z) = -1/[z-1/m_q(z)]$ \cite{Goetze1992}.
Equivalently, there holds $\phi_q(z)/[1-z\phi_q(z)] = m_q(z)$. It will be
more convenient to modify the Laplace transform to another invertible
mapping ${\cal S}$ from the time domain to the domain of complex
frequencies according to

\begin{equation}\label{eq:def_S}
{\cal S}[F(t)](z) = (-iz)\int_0^\infty\,dt\,\exp(izt)\,F(t)\,.
\end{equation}
Using this notation, the MCT equations for the small-frequency dynamics read
\begin{equation}\label{eq:MCT}
{\cal S}[\phi_q(t)](z)/\{1-{\cal S}[\phi_q(t)](z)\} = {\cal S}\left[{\cal 
F}_q[\mathbf{V},\phi_k(t)]\right](z)\,.
\end{equation}
Since ${\cal F}_q$ is determined completely by the equilibrium structure
functions, the dynamics obtained from Eqs.~(\ref{eq:def_m})
and~(\ref{eq:MCT}) is referred to as structural relaxation. These
equations are scale invariant: if $\phi_q(t)$ is a solution, the same is
true for $\phi^x_q(t) = \phi_q(x t)$ for all $x > 0$. The scale for the
dynamics is determined by the transient motion.  The latter is governed by
$\Omega_q$ and $M^{\rm reg}_q(t)$. Since these quantities do not enter
Eq.~(\ref{eq:MCT}), the solutions of Eqs.~(\ref{eq:def_m}) and~(\ref{eq:MCT}) 
are fixed only up to an overall time scale \cite{Goetze1992}. In the 
following, this time scale will be denoted by $t_0$.

A glass state is characterized by non-vanishing long-time limits of the
correlators:  $\lim_{t\rightarrow\infty}\phi_q(t) = f_q,\, 0<f_q<1$. 
Equivalently, one gets $\lim_{z\rightarrow 0}{\cal S}[\phi_q(t)](z) = f_q$. 
Hence, the zero-frequency limit of Eq.~(\ref{eq:MCT})  yields $f_q/(1-f_q) = 
{\cal F}_q[\mathbf{V},f_k]$, $q=1,\,2,\,\dots,\,M$. 
This is a set of $M$ implicit equations to be
obeyed by the $ M$ numbers $f_q$ \cite{Bengtzelius1984}. If the Jacobian
of these equations is invertible, the solutions vary smoothly with changes
of $\mathbf{V}$. If the Jacobian is singular for some state $\mathbf{V}^c$
with $f_q= f_q^c$, $f_q$ exhibits a singularity as function of
$\mathbf{V}$ for $\mathbf{V}$ tending towards $\mathbf{V}^c$. Therefore,
such state $\mathbf{V}^c$ is called glass-transition singularity. The
solution for the correlators for $\mathbf{V}=\mathbf{V}^c$ is referred to
as critical correlator $\phi^c_q(t)$. Let us introduce a functions
$\hat{\phi}_q(t)$ by

\begin{equation}\label{eq:def_hatphi}
\phi^c_q(t) = f_q^c + (1-f_q^c) \hat{\phi}_q(t)
\end{equation}
obeying $\lim_{t\rightarrow\infty}\hat{\phi}_q(t) = 0$. In the following,
$\hat{\phi}_q(t)$ and ${\cal S}[\hat{\phi}_q(t)](z)$ shall be used as
small quantities for an asymptotic expansion of $\phi^c_q(t)$ for large
times and small frequencies. Introducing the coefficients
\begin{equation}\label{eq:def_Acoeff}\fl\quad
A^{(n)c}_{qk_1\dots k_n} = \frac{1}{n!} (1-f_q^c) \{
\partial^n {\cal F}_q[\mathbf{V}^c,f_k^c]/\partial f_{k_1}\cdots\partial 
f_{k_n}
\}
(1-f_{k_1}^c)\cdots(1-f_{k_n}^c) \,,
\end{equation}
Eq.~(\ref{eq:MCT}) can be rewritten as the set of equations of motion for 
the $\hat{\phi}_q(t)$ \cite{Goetze1991b}:
\begin{equation}\label{eq:eom_hatphi}
[\delta_{qk}-A^{(1)c}_{qk}]\, {\cal S}[\hat{\phi}_k(t)](z) = J_q(z)\,.
\end{equation}
Here, $J_q(z) = \sum_{n\geq 2} J_q^{(n)}(z)$ with the $n$th order expansion 
term given by
\begin{equation}\label{eq:def_Jqn}
J_q^{(n)}(z) = A^{(n)c}_{qk_1\dots k_n} {\cal 
S}[\hat{\phi}_{k_1}(t)\cdots\hat{\phi}_{k_n}(t)](z)
-  {\cal S}[\hat{\phi}_q(t)]^n(z)
\,.
\end{equation}
In Eq.~(\ref{eq:eom_hatphi}) and ~(\ref{eq:def_Jqn}) and in all the following 
equations, summation over pairs of equal labels 
$k$ is implied. The $M\times M$ matrix $[\delta_{qk}-A^{(1)c}_{qk}]$ is the 
Jacobian mentioned above. Therefore,
a singularity is characterized by matrix $A^{(1)c}_{qk}$ to have an
eigenvalue unity. It is a subtle property of the MCT equations that this
eigenvalue is non-degenerate and that all other eigenvalues of
$A^{(1)c}_{qk}$ have a modulus smaller than unity. The left and right 
eigenvectors shall be denoted by $a^*_q$ and $a_q,\,q=1,\dots,M$, respectively,

\begin{equation}\label{eq:eigenvector}
a^*_k A^{(1)c}_{kq} = a^*_q \,,\quad A^{(1)c}_{qk} a_k = a_q
\,.
\end{equation}
Generically, one can require $a_q\geqslant 0$, $a_q^*\geqslant 0$ for 
$q=1,\dots,M$. To fix the eigenvectors uniquely, two normalization conditions 
can be imposed: $\sum_q a^*_q a_q =1,\, \sum_q a^*_q a_q a_ q= 1$
\cite{Goetze1991b,Goetze1995b}.

Because of the non-degeneracy mentioned, the singularity is topologically
equivalent to that of the zeros of a real polynomial of degree $l,\,l=2,\,3,
\dots$. It is a bifurcation of type $A_l$ \cite{Arnold1986}. The singularity 
can be characterized by a sequence of real coefficients $\mu_2,\,\mu_3,\,
\dots$. An $A_l$ is specified by $\mu_n =0$ for $n<l$ and $\mu_l \neq 0$. 
The simplest of these numbers reads

\begin{equation}\label{eq:mu2}
\mu_2 = 1 - \sum_q a^*_q  A^{(2)c}_{qk_1k_2} a_{k_1} a_{k_2}\,.
\end{equation}
For an $A_2$-glass-transition singularity, $\mu_2$ determines the
so-called critical exponent $a,\,0<a\leq 1/2$. In this case, the critical
correlator can be asymptotically expanded as a power series:  
$\hat{\phi}_q(t) = a_q (t_0/t)^a + a'_q (t_0/t)^{2a} +\cdots$ . If the
$A_2$ singularity approaches a higher-order singularity $A_l,\,l\geq 3$,
the exponent $a$ approaches zero and the cited asymptotic expansion breaks
down \cite{Goetze1991b}. It is the goal of this paper to derive a
long-time expansion of the critical correlator at $A_3$- and 
$A_4$-singularities. Equivalently, it is the aim to solve asymptotically
Eqs.~(\ref{eq:eom_hatphi})  and~(\ref{eq:def_Jqn}) for $\hat{\phi}_q(t)$
for states $\mathbf{V}^c$ with
\numparts
\begin{equation}\label{eq:def_A3}
\mu_2 = 0\,,\quad\mu_3\neq 0\,
\end{equation}
for an $A_3$-singularity denoted by $\mathbf{V}=\mathbf{V}^\circ$ and
\begin{equation}\label{eq:def_A4}
\mu_2 = \mu_3 = 0\,,\quad\mu_4\neq 0\,
\end{equation}
for an $A_4$-singularity denoted by $\mathbf{V}=\mathbf{V}^*$.
\endnumparts

\subsection{\label{sec:expansion}Expansions of slowly-varying functions}

The derivations in this paper shall be based on an extension of the Tauberian
theorem for slowly-varying functions, which has been
introduced in Ref.~\cite{Goetze1989d}.  A function $C(t)$ is called of
slow variation for long times if $\lim_{T\rightarrow\infty} C(tT)/C(T) =
1$ for all $t>0$. This is equivalent to $\gamma(z) = { \cal S}[C(t)](z)$
being slowly varying for small frequencies: $\lim_{T\rightarrow\infty}
\gamma(z/T)/\gamma(i/T) = 1$. In addition, the Tauberian theorem states,
that $\gamma(z)$ is asymptotically equal to $G(i/z):\,\lim_{z\rightarrow
0} \gamma(z)/ G(i/z) = 1$ \cite{Feller1971b}. Typical examples for functions of
slow variation are $p_m(\ln(\ln t))/\ln^m(t)$, where $m=1,\,2,\,\dots$ and 
$p_m$ denotes some polynomial. The critical correlator $\phi^c_q(t)$ shall be 
expressed as sum of such functions. Let us introduce the notations
\numparts\label{eq:logpoly}
\begin{equation}\label{eq:logpoly_a}
G(t) = g(x)\,,\quad
x = \ln (t/t_0)
\,,\quad
y = \ln(i/z t_0)\,,
\end{equation}
\begin{equation}\label{eq:logpoly_b}
g_m(x) = p_m(x)/x^m
\,,\quad
p_m(x) = \sum_{l=0}^{l_0} c_{m,l} x^l\,.
\end{equation}
\endnumparts
$g_{m+1}(x)$ is asymptotically negligible compared to
$g_{m}(x)$: $\lim_{x\rightarrow\infty} g_{m+1}(x) / g_{m}(x) = 0$. For
later convenience, let us write $f(x) = {\cal O}(1/x^m)$ if $f(x) x^m$ is
bounded for large $x$ by some polynomial of $\ln x$. Denoting derivatives
by $d^n g(x)/dx^n = g^{(n)}(x),\,n=0,\,1,\,\dots$, one finds

\begin{equation}\label{eq:logpoly_derivative}
g_m^{(n)}(x) = {\cal O}(1/x^{m+n})\,.
\end{equation}

Equation~(\ref{eq:def_S}) can be rewritten as ${\cal S}[G(t)](z) =
\int_0^\infty\,\exp(-u) g(y+\ln u) \,du$. Formal expansion in powers of
$\ln u$ leads to
\begin{equation}\label{eq:polylog_expansion}
{\cal S}[G(t)](z) = \sum_{n=0}^\infty \frac{1}{n!} \Gamma_n g^{(n)}(y)\,.
\end{equation}
Here, $\Gamma_n = \Gamma^{(n)}(1)$ denotes the $n$th derivative of the
gamma function at unity. One gets $\Gamma_0 = 1,\,-\Gamma_1 = \gamma$ is
Euler's constant, and $\Gamma_n$ for $n\geq 2$ can be expressed in terms
of $\gamma$ and Riemann's zeta-function values $\zeta(K)$ with
$K=2,\,\dots,\,n$ \cite{Abramowitz1970}. For example, $\Gamma_2 -
\Gamma_1^2 = \zeta(2)$. Using Eq.~(\ref{eq:polylog_expansion}) with $G(t)
= g_m(x)$, one gets an asymptotic expansion in terms of increasing order
${\cal O}(1/y^{m+n})$. The leading $n=0$ contribution is $g_m(y)$; and
this is the result of the Tauberian theorem \cite{Feller1971b}. The
terms for $n\geq 1$ provide systematic improvements for large $y$, i.e.
for large times or small frequencies \cite{Goetze1989d}.

If one uses Eq.~(\ref{eq:polylog_expansion}) for $G(t) = G(t) F(t)$, one 
gets the asymptotic expansion
\begin{equation}\label{eq:expansion_GF}
\fl
{\cal S}[G(t) F(t)](z) - {\cal S}[G(t)](z) {\cal S}[F(t)](z)  =
\sum_{n=2}^\infty \sum_{m=1}^{n-1}\frac{[\Gamma_n - 
\Gamma_{n-m}\Gamma_n]}{(n-m)!m!}  g^{(n-m)}(y) f^{(m)}(y).
\end{equation}
Let us use $G(t)=g_{m_1}(x)$ and $F(t)=g_{m_2}(x)$. The Tauberian theorem
implies that the leading contribution to ${\cal S}[G_{m_1}(t)
G_{m_2}(t)](z)$ cancels against the leading contribution to ${\cal
S}[G_{m_1}(t)](z)  {\cal S}[G_{m_2}(t)](z)$. The tricks underlying the
asymptotic solution of the MCT equations at a higher-order singularity are
based on the observation that also the leading corrections to the
Tauberian theorem cancel \cite{Goetze1989d}:

\begin{equation}\label{eq:Tauber_cancellation}\fl\qquad
{\cal S}[g_{m_1}(t) g_{m_2}(t)](z) - {\cal S}[g_{m_1}(t)](z) {\cal 
S}[g_{m_2}(t)](z)
 = {\cal O}(1/y^{m_1+m_2+2})\,.
\end{equation}
The difference between the two terms on the left-hand side is two orders
smaller for vanishing frequencies than each of the terms separately.

\section{\label{sec:crit_one_A3}Critical correlators for one-component 
models at an $A_3$-singularity}

\subsection{\label{sec:leading_one_A3}The leading contribution}

It will be shown in Sec.~\ref{sec:critA3} how one can reduce the problem
of solving Eqs.~(\ref{eq:eom_hatphi})  and~(\ref{eq:def_Jqn}) for a
general number $M$ of the correlators to the special problem of solving
for $M=1$ models. Therefore, the problem shall be discussed first for
one-component models.  For this case, one can drop the indices in all formulas 
of Sec.~\ref{sec:A3A4}. There is only one
correlator $\phi^c(t)$, one long-time limit $f^c$ for the critical point
$\mathbf{V}^c$, and one function $\hat{\phi}(t)$ determining the
critical correlator as $\phi^c(t) = f^c + (1-f^c) \hat{\phi}(t)$. The
Jacobian matrix agrees with its eigenvalue, and this is zero. Hence,
Eqs.~(\ref{eq:eom_hatphi}) and~(\ref{eq:def_Jqn}) can be noted as
\numparts\label{eq:def_K}
\begin{equation}\label{eq:def_K_a}
K(z) = 0 \,,
\end{equation}
\begin{equation}\label{eq:def_K_b}
K(z) = \sum_{n=2}^\infty K_n(z)\,.
\end{equation}
\endnumparts
Here, $K_n(z)$ is the expansion term of order $\hat{\phi}^n$. Let us
introduce the abbreviation
\begin{equation}\label{eq:def_psi}
\psi_n(z) =  {\cal S}[\hat{\phi}^n(t)](z)  - {\cal S}[\hat{\phi}(t)]^n(z) 
\,,
\end{equation}
and denote its inverse transform by $\psi_n(t)$, i.e., ${\cal S}[\psi_n(t)](z) 
= \psi_n(z)$. Remembering that for $M=1$ models there holds $\mu_n=1-A^{(n)c}$,
one gets $K_n(z)=\psi_n(z)-\mu_n{\cal S}[\hat{\phi}^n(t)](z)$ 
\cite{Goetze2002}.  Specializing to the $A_3$-singularity as noted in 
Eq.~(\ref{eq:def_A3}), the equation of motion~(\ref{eq:def_K_a}) is defined by
\begin{equation}\label{eq:eom_K} 
\begin{array}{ll}
  K(z) = & \psi_2(z) - \mu_3 {\cal S}[\hat{\phi}^3(t)](z)\\
         & + \kappa \psi_3(z) - \mu_4 {\cal S}[\hat{\phi}^4(t)](z)\\
         & + K'(z)
\,.
\end{array}
\end{equation}
Here, $K'(z) = \kappa'\psi_4(z) - \mu_5 {\cal S}[\hat{\phi}^5(t)](z)+\dots$. 
The numbers $\kappa$ and $\kappa'$ have been introduced for later convenience.
For the $M=1$ models under consideration, one has to substitute 
$\kappa=\kappa'=1$.

Let us examine whether one can solve the equations with the Ansatz
$\hat{\phi}(t) = g_m(x) = c_m/x^m$. From
Eq.~(\ref{eq:polylog_expansion}) one gets ${\cal S}[\hat{\phi}^3(t)](z) =
(c_m/y^m)^3 + {\cal O}(1/y^{3m+1})$. Using Eq.~(\ref{eq:expansion_GF}) with
$G(t) = F(t) = g_m(x)$, one obtains $\psi_2(z) = \zeta(2)(m c_m/y^{m+1})^2
+ {\cal O}(1/y^{2m+3})$. Choosing $m=2$, both terms in the first line of
Eq.~(\ref{eq:eom_K}) are of the same order $1/y^6$. They cancel in this
leading order if $\mu_3 c_2^3 = 4 \zeta(2)  c_2^2$. From
Eqs.~(\ref{eq:polylog_expansion})  and~(\ref{eq:Tauber_cancellation}) one
infers that the terms in the second line of Eq.~(\ref{eq:eom_K}) are of
order $1/y^8$ and $K' = {\cal O}(1/y^{10})$. One concludes that the
leading asymptotic behavior of the critical correlator for large times is
described by $\hat{\phi}(t) = g_2(x)$, where
\begin{equation}\label{eq:g2} 
g_2(x) = c_2/x^2\,,\quad c_2 =  4 \zeta(2) / \mu_3\,.
\end{equation}

\subsection{\label{sec:correction_one_A3}The leading correction}

Let us split the function $\hat{\phi}(t)$ into its leading term and a
correction $\tilde{g}(x)$:
\begin{equation}\label{eq:tildeg} 
\hat{\phi}(t) = g_2(x) + \tilde{g}(x)\,.
\end{equation}
Substitution of this formula into the first line of Eq.~(\ref{eq:eom_K}),
one gets expressions up to third order in $\tilde{g}$. The term
independent of $\tilde{g}$ is ${\cal S}[g_2^2(x)](z)-{\cal
S}[g_2(x)]^2(z)-\mu_3{\cal S}[g_2^3(x)](z)$, and it shall be denoted by
$[(4 \zeta(2))^2 / \mu_3] F(y)$.  Equations~(\ref{eq:polylog_expansion})
and~(\ref{eq:expansion_GF}) are used to derive the asymptotic series
\numparts\label{eq:corrections}
\begin{eqnarray}\label{eq:Fy}\fl\qquad
F(y) = \sum_{n=3}^\infty \frac{(-1)^{n+1}}{\mu_3 y^{4+n}}
  \left\{\frac{1}{30}\zeta(2) \frac{(n+3)!}{(n-2)!} \Gamma_{n-2}\right.
\nonumber\\ -
 \left.   \sum_{m=1}^{n-2} (n-m+1)(m+1)(\Gamma_n-\Gamma_{n-m}\Gamma_m)\right\}.
\end{eqnarray}
The term linear in $\tilde{g}$ is $2\{{\cal 
S}[g_2(x)\tilde{g}(x)](z)-{\cal S}[g_2(x)](z){\cal S}[\tilde{g}(x)](z)\}-
3 \mu_3{\cal S}[g_2^2(x)\tilde{g}(x)](z)$. 
It shall be denoted by $[(4 \zeta(2))^2 / \mu_3][{\cal D}\tilde{g}(y) + 
{\cal D}'\tilde{g}(y)]$. Here, the differential 
operator ${\cal D}$ yields the leading contribution
\begin{equation}\label{eq:def_D} 
{\cal D}\tilde{g}(y) = [y\cdot d\tilde{g}(y)/dy+ 3 \tilde{g}(y)]/y^4\,.
\end{equation}
The correction ${\cal D}'$ is expanded with the aid of 
Eqs.~(\ref{eq:polylog_expansion}) and~(\ref{eq:expansion_GF}):
\begin{eqnarray}\label{eq:def_Ddash}\fl\quad
{\cal D}'\tilde{g}(y) = [1/2 \zeta(2)] \sum_{n=3}^\infty 
\sum_{m=1}^{n-1}(-1)^{n-m}
\left\{[\tilde{g}^{(m)}(y)/y^{n+2-m} m!] (\Gamma_n-\Gamma_{n-m}\Gamma_m)
\right.\nonumber\\\fl\qquad\left.
 + \zeta(2) \Gamma_{n-2} [\tilde{g}^{(m-1)}(y)/y^{n+3-m} (m-1)!] 
(n-m+1)(n-m)(n-m+1)
\right\}
\,.
\end{eqnarray}
\endnumparts
With these notations, the equation of motion for $\tilde{g}(y)$ is
reformulated as a linear differential equation with some
inhomogeneity $I(y)$:
\numparts
\label{eq:ODE}
\begin{equation}\label{eq:ODE:Dg} 
{\cal D}\tilde{g}(y) = I(y)\,,
\end{equation}
\begin{eqnarray}\label{eq:ODE:I}
I(y) =& F(y) + {\cal D}'\tilde{g}(y)\nonumber\\
      &+ {\cal S}[\tilde{g}^2(x)](z) - {\cal S}[\tilde{g}(x)]^2(z) - 3 
\mu_3{\cal S}[g_2(x)\tilde{g}^2(x)](z)\nonumber\\
      & -\mu_3{\cal S}[\tilde{g}^3(x)](z)+\kappa\psi_3(z) -\mu_4 {\cal 
S}[\hat{\phi}^4(t)](z) + K'(z)
\,.\end{eqnarray}
\endnumparts
It might be adequate to emphasize, that
Eqs.~(\ref{eq:g2})--(\ref{eq:ODE:I}) formulate an exact rewriting of
Eq.~(\ref{eq:MCT}) for $M=1$ models.

The iterative solution of Eq.~(\ref{eq:ODE:Dg} ) for $\tilde{g}(x)$ is
based on the observation, that one gets for functions $g_m(y)$ from
Eq.~(\ref{eq:logpoly_b}):
\begin{equation}\label{eq:ODE_iter}
{\cal D}g_m(y) = [p_m'(y) + (3-m)p_m]/y^{m+4}\,.
\end{equation}
If one tries with $\tilde{g}(x) = g_3(x)$, one finds on the one hand
${\cal D}g_m(y) = p_3'(y)/y^7$. On the other hand, one verifies, that all
terms on the right hand side of Eq.~(\ref{eq:ODE:I}) are ${\cal O}(1/y^8)$
except for the $n=4$ contribution to $F(y)$. One checks that $F(y) = 24
\zeta(3)/(\mu_3 y^7) + {\cal O}(1/y^8)$. Hence, the leading order solution
for $\tilde{g}$ reads

\begin{equation}\label{eq:g3}
g_3(x) = c_3 \ln(x)/x^3\,,\quad c_3 = 24 \zeta(3)/\mu_3\,.
\end{equation}
Combining this finding with Eqs.~(\ref{eq:g2}) and~(\ref{eq:tildeg}) and
eliminating all the abbreviations, one reproduces a result of
Ref.~\cite{Goetze1989d}:
\begin{equation}\label{eq:GSj_result}\fl\qquad
\phi^\circ(t) = f^\circ + (1-f^\circ)[c_2/\ln^2(t/t_0)] 
\left\{1+[6\zeta(3)/\zeta(2)]\ln\ln(t/t_0)/\ln(t/t_0)\right\} \,.
\end{equation}
This formula describes the critical correlator up to errors of the order
$1/\ln^4(t/t_0)$.

\subsection{\label{sec:highcorrection_one_A3}Higher-order contributions}

The equation for $\tilde{g}(y)$ allows for an iterative solution so that
the iteration step with number $m$ reads $\tilde{g} = g_3 + g_4 + \cdots +
g_m$. Here the numerator polynomial in Eq.~(\ref{eq:logpoly_b}) is of
degree not larger than $(m-2)$, i.e.,

\begin{equation}\label{eq:gm} 
g_m(x) = \sum_{l=0}^{m-2} c_{m,l} \ln^l(x)/x^m \,.
\end{equation}
Suppose, the procedure had been carried out up to step $m-1,\,m=4,\,5,
\dots$. Then ${\cal D}\tilde{g}(y) = {\cal D}g_m(y) + {\cal
O}(1/y^{m+3})$. By construction, all terms up to order $(m+3)$ cancel
against the one appearing in $I(y)$. One checks, that the leading
contribution to $I(y)$ reads $p(\ln y)/y^{m+4}$, where the degree of the
polynomial $p$ does not exceed $m-3$. Hence, Eq.~(\ref{eq:ODE:Dg}) is
equivalent to the linear differential equation $p_m'+(3-m)p_m=p$. It is
readily solved by Eq.~(\ref{eq:gm}), provided the coefficients $c_{m,l}$
are chosen properly.

In order to determine $g_4$ and $g_5$, one can drop $K'(z)$ in
Eq.~(\ref{eq:ODE:I}). The coefficients $c_{m,l}$ are given by $\mu_3$,
$\kappa$, and $\mu_4$ as follows
\numparts\label{eq:c4}
\begin{eqnarray}\label{eq:c40}\fl\qquad
c_{4,0} &=& 792 \zeta(3)^2/(\pi^2\mu_3) + 
[4\mu_4/(9\mu_3^2)-4\kappa/(3\mu_3)-7/6]\,\pi^4/\mu_3\,,
\\\fl\qquad\label{eq:c41}
c_{4,1} &=&  -432 \zeta(3)^2/(\pi^2\mu_3)\,,
\\\fl\qquad\label{eq:c42}
c_{4,2} &=&  648 \zeta(3)^2/(\pi^2\mu_3)\,,
\end{eqnarray}
\endnumparts
\numparts\label{eq:c5}
\begin{eqnarray}\label{eq:c50}\fl\qquad
c_{5,0} &=& \zeta(3)\pi^2[400 \kappa\mu_3 + 1551\mu_3^2 -160 
\mu_4]/(15\mu_3^3)\nonumber\\\fl\qquad
 &&-[39744\,\zeta(3)^3/\pi^4 + 528\zeta(5)]/\mu_3\,,
\\\fl\qquad\label{eq:c51}
c_{5,1} &=& 64800\,\zeta(3)^3/(\pi^4\mu_3)-4\zeta(3)\pi^2(21 \mu_3^2 
-24\kappa\mu_3 +8\mu_4)/\mu_3^3\,,
\\\fl\qquad\label{eq:c52}
c_{5,2} &=& -27216\,\zeta(3)^3/(\pi^4\mu_3)\,,
\\\fl\qquad\label{eq:c53}
c_{5,3} &=& 15552\,\zeta(3)^3/(\pi^4\mu_3)\,,
\end{eqnarray}
\endnumparts
The coefficients for $g_6$ and $g_7$ have also been determined . The only new
model parameters entering the coefficients are $\mu_5$ and $\kappa'$ 
\cite{Sperl2003}.

\subsection{\label{sec:discussion_one_A3}Discussion}

The preceding results shall be demonstrated quantitatively for the simplest 
model exhibiting a generic $A_3$-glass-transition singularity. This model was
derived for a spin-glass system and it is defined by the mode-coupling function
\cite{Goetze1984b}
\begin{equation}\label{eq:F13} 
m(t) = v_1 \phi(t) + v_3 \phi^3(t) \,.
\end{equation}
Here and in the following models, we use a Brownian short-time dynamics as
specified by the equation of motion
\begin{equation}\label{eq:shorttime:Brown} 
\tau\partial_t \phi(t) + \phi(t) + \int_0^t\, dt' m(t-t') 
\partial'_t \phi(t') = 0 \,,
\end{equation}
to be solved with the initial condition $\phi(t\rightarrow 0)=1$. The 
short-time asymptote is $\phi(t)-1=-(t/\tau)+{\cal O}((t/\tau)^2)$.
The singularity is obtained for the coupling constants $v_1^\circ = 9/8$ and 
$v_3^\circ = 27/8$. The critical long-time limit of the correlator is 
$f^\circ=1/3$ \cite{Goetze2002,Goetze1988b}. The other parameters entering the 
coefficients via Eqs.~(\ref{eq:c4}) and~(\ref{eq:c5}) are $\mu_3=1/3$ and 
$\mu_4=\mu_5=\kappa=\kappa'=1$. Thus, all expansion formulas are specified, 
except for the time scale $t_0$. To ease reference to various degrees
of asymptotic expansions, let us introduce the abbreviation for the $n$th order
approximation
\begin{equation}\label{eq:sol_n} 
\phi^\circ(t)_n = f^\circ + (1-f^\circ)\,G_n(t)\,,\quad
G_n(t) = \sum_{m=2}^{n} g_m(\ln(t/t_0)) \,.
\end{equation}

\begin{figure}[htb]
\includegraphics[width=\columnwidth]{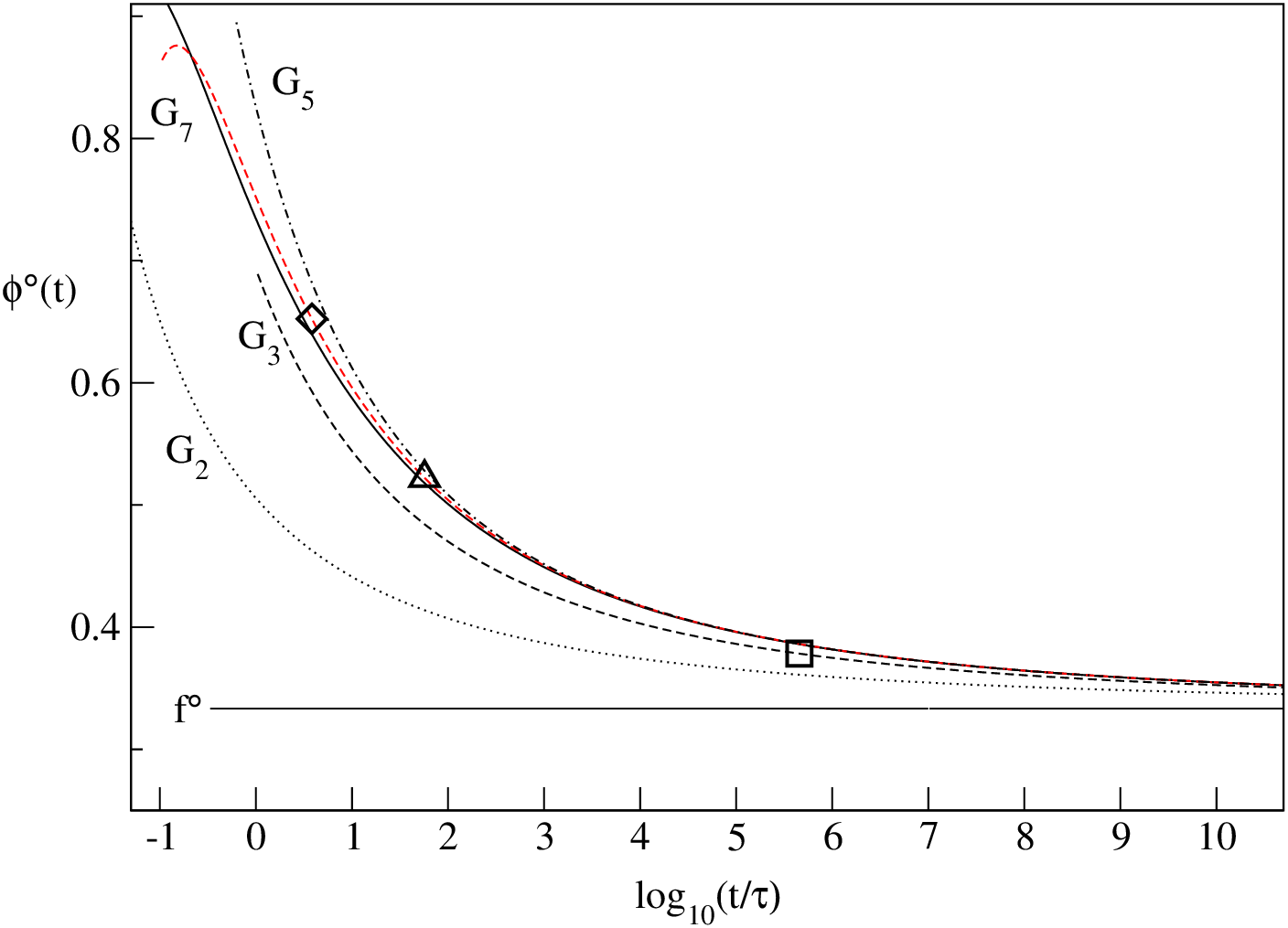}
\caption{\label{fig:F13}Critical decay at the $A_3$-singularity of the model
defined by Eqs.~(\ref{eq:F13}) and~(\ref{eq:shorttime:Brown}). The full line
shows the solution for $\phi^\circ(t)$. The lines labeled $G_n$, for 
$n=2,\,3,\,5,\,7$, show the approximations from Eq.~(\ref{eq:sol_n}) with the 
time scale $t_0/\tau=1.6\cdot10^{-4}$. The time where $G_3$, $G_5$, and $G_7$ 
deviate by $2\%$ from $\phi^\circ(t)$ is marked by a square ($\square$), a 
triangle ($\triangle$), and a diamond ($\diamond$), respectively.
}
\end{figure}
Figure~\ref{fig:F13} exhibits $\phi^\circ(t)$ as obtained from 
Eqs.~(\ref{eq:F13}) and~(\ref{eq:shorttime:Brown})
for the state $\mathbf{V}=\mathbf{V}^\circ$.
The approach to the critical plateau $f^\circ$ is significantly slower than 
the one for a typical $A_2$-singularity. In the latter case, the decay comes
close to the plateau within a few decades of increase of the time when a 
deviation of $5\%$ is used as a measure. Such a criterion is not met by the
decay in Fig.~\ref{fig:F13} for the entire window in time shown. For 
$t=10^{11}$, the critical correlator $\phi^\circ(t)$ is still
$5.5\%$ above $f^\circ$. To apply the asymptotic approximation, one 
has to match the time scale $t_0$ at large times. A reliable determination of 
$t_0$ is not possible when using only $G_2(x)$ or $G_3(x)$. Using $G_7(x)$ and 
extending the numerical solution to $t/\tau=10^{38}$, it is possible to fix 
$t_0/\tau=1.6\cdot10^{-4}$. Notice, that 
$t_0$ is several orders of magnitude smaller than the time scale $\tau$ for
the transient dynamics. With this value for $t_0$, the successive asymptotic 
approximations are shown in Fig.~\ref{fig:F13}. The leading approximation
from Eq.~(\ref{eq:sol_n}), labeled $G_2$, deviates from the critical correlator
strongly. Including the next-to-leading term $g_3(x)$ yields the approximation 
labeled $G_3$, i.e. Eq.~(\ref{eq:GSj_result}). A square indicates that $G_3$ 
deviates from the critical correlator by less than $2\%$ for $t/\tau\gtrsim 
5\cdot10^5$. If that criterion is relaxed to $5\%$, $G_3$ obeys it for 
$t\gtrsim 10^3\tau$. The
approximation by $G_3$ provides a first reasonable approximation to 
$\phi^\circ(t)$. Including further terms of the expansion improves the 
approximation as is shown for $G_5$ and $G_7$. One recognizes that proceeding
from $G_5$ to $G_7$
still improves the range of applicability by one order of magnitude in time.
We conclude that the asymptotic expansion explains quantitatively the critical 
decay at the $A_3$-singularity for all times outside the transient regime.

\begin{figure}[htb]\includegraphics[width=0.9\columnwidth]{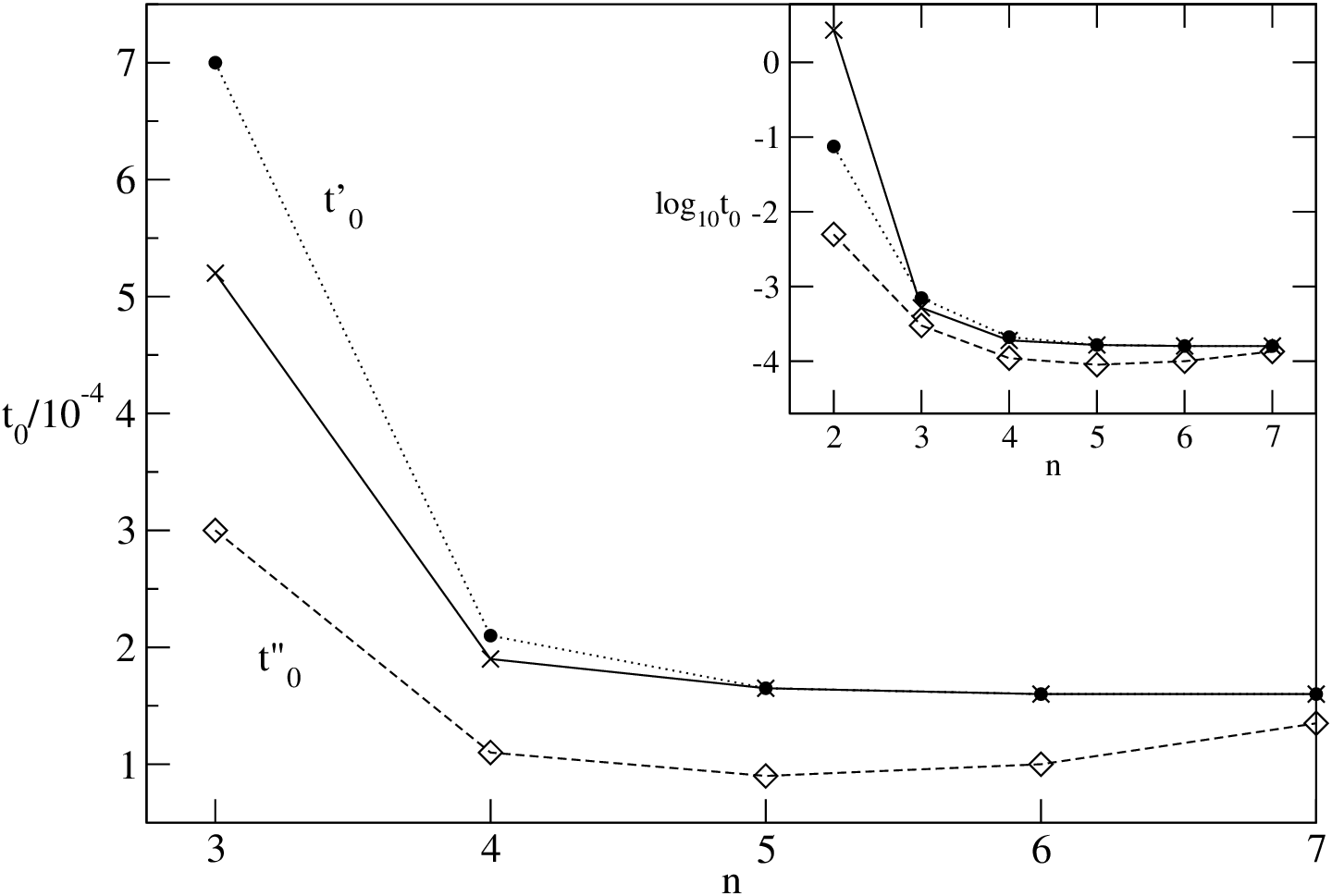}
\caption{\label{fig:F13t0}Time scale $t_0$ in units of $\tau$ for the 
approximation of the critical decay at the $A_3$-singularity of the model
studied in Fig.~\ref{fig:F13} by including $n$ orders of the asymptotic 
expansion, Eq.~(\ref{eq:sol_n}). Time scales obtained by matching $G_7(t)$ at 
large time, $35\lesssim\log_{10}t\lesssim 38$ are shown 
by crosses ($\times$). The time $t'_0$ resulting from matching the 
solutions at $t=10^6$ is shown by filled circles ($\bullet$). The diamonds 
($\diamond$) show the time scale $t_0''$ resulting from matching
where $\phi(t)=2/3$. The inset shows $t_0$ on logarithmic scale. The
lines are guide to the eye.
}
\end{figure}
Matching a time scale $t_0$ at $t/\tau=10^{40}$ and using six 
terms of the expansion in Eq.~(\ref{eq:sol_n}) is not a promising perspective
for fitting data. However, the expansion leads to a reasonable approximation
also for short times. Therefore, we may depart from the procedure to match 
$t_0$ at large
times and try to fit $t_0$ for shorter times. Figure~\ref{fig:F13t0} shows as 
crosses the values obtained for $t_0$ when matching the approximations at the 
large times mentioned above. We will
consider two procedures for fitting. The first shall define a scale $t_0'$
by matching the critical correlator by the approximation at $t=10^6$.
The second time scale $t_0''$ is obtained from matching at $50\%$ of the decay,
i.e. for the time $t^*$ where $\phi^\circ(t^*)=2/3$.
We infer from the inset of Fig.~\ref{fig:F13t0} that all methods
to fix $t_0$ based on the term $G_2(x)$ alone are off by orders of
magnitude. The approximation $G_3(x)$ yields the correct order of
magnitude for $t_0$ in all three approaches. Starting with $n=5$, the
scales $t_0$ and $t'_0$ can no longer be distinguished. Therefore, matching
the approximation at $10^6$ is comparable to matching a true asymptotic limit.
The value $t_0'$ is a better approximation for $t_0$ than $t_0''$.

\section{\label{sec:crit_one_A4}Critical correlators for one-component 
models at an $A_4$-singularity}

Within the theory of the logarithmic decay as presented in 
Ref.~\cite{Goetze2002}, it is possible to specialize to the $A_4$-singularity
by simply setting $\mu_3=0$ in the final formulas. Different from that, the 
critical decay for the $A_4$-singularity does not follow from the solution for
the $A_3$-singularity but requires a different asymptotic expansion. This can 
be inferred from the fact that all the coefficients $c_{m,l}$ in 
Eq.~(\ref{eq:gm}) contain $\mu_3$ in the denominator. However, the tricks used
for finding a solution in terms of slowly varying functions are the same for 
the $A_4$ as explained above for the $A_3$.

\subsection{\label{sec:leading_one_A4}The leading contribution}

Using Eq.~(\ref{eq:def_A4}) for an $A_4$-singularity, 
Eqs.~(\ref{eq:def_K_a}) and~(\ref{eq:eom_K}) can be regrouped as 
\begin{equation}\label{eq:problemA4}
\begin{array}{cccc}
0 =\qquad
&\psi_2(z)
        &-& \mu_4\,{\cal S}[\hat{\phi}^4(t)](z)\\
\quad+&\kappa\psi_3(z)
        &- &\mu_5\,{\cal S}[\hat{\phi}^5(t)](z)\\
\quad+&\kappa'\psi_4(z)
        &- &\mu_6\,{\cal S}[\hat{\phi}^6(t)](z)\\
&&&+\qquad\dots\,\,.
\end{array}
\end{equation}
With the Ansatz $\hat{\phi}(t) = g_m(x)= c_m/x^m$, one arrives for the terms of
the first line at $\psi_2(z)=\zeta(2)(mc_m/y^{m+1})^2+{\cal O}(1/y^{2m+3})$ and
${\cal S}[\hat{\phi}^4(t)](z)=(c_m/y^m)^4+{\cal O}(1/y^{4m+1})$. For $m=1$, the
first line in Eq.~(\ref{eq:problemA4}) is of leading order ${\cal O}(1/y^4)$ 
with the equation for the coefficient $\zeta(2)c_1^2=\mu_4c_1^4$. This results 
in the leading-order solution 
\cite{Goetze1989d},
\begin{equation}\label{eq:A4g1}
g_1(x) = c_1/x\,,\quad c_1 =  \sqrt{ \zeta(2) / \mu_4}\,.
\end{equation}

\subsection{\label{sec:correction_one_A4}The leading correction}

The corrections may be rephrased in terms of a differential operator and the 
solution is straight forward as before. Since later on, only the first 
correction will be needed explicitly, it will be calculated here by the linear 
differential equation for the Ansatz 
$\hat{\phi}(t)=\left[\phi^*(t)-f^*\right]/(1-f^*)=g_1(x)+\tilde{g}(x)$,
\begin{equation}\label{eq:A4crit:corr1}\fl\qquad
2 y^3 \tilde{g}'(y) + 4 y^2 \tilde{g} 
        =4\sqrt{\zeta(2)/\mu_4}\,\zeta(3)/\zeta(2)
                +3\zeta(2)\kappa/\mu_4 -\mu_5\zeta(2)/\mu_4^2\,.
\end{equation}
This is solved in leading order by $g_2(x)$:
\begin{equation}\label{eq:A4g2}\fl
g_2(x) = c_2\ln(x)/x^2\,,\quad c_2 =
        2\sqrt{\zeta(2)/\mu_4}\,\zeta(3)/\zeta(2)
          + 3\zeta(2)\kappa/(2\mu_4)
                -\mu_5\zeta(2)/(2\mu_4^2)\,.
\end{equation}
Higher-order contributions for $m\geqslant 3$ can be written in the form
\begin{equation}\label{eq:A4:gm} 
g_m(x) = \sum_{l=0}^{m-1} c_{m,l} \ln^l(x)/x^m
\end{equation}
with the appropriate choice of the parameters $c_{m,l}$. Hence, the 
general solution for the critical decay at an $A_4$-singularity in the 
one-component case is represented up to errors of order
${\cal O}(\ln^{-(n+1)} (t)$ as 
\begin{equation}\label{eq:A4sol}
\phi(t)^*=f^*+(1-f^*)\, G_n(t)\,,\quad
G_n(t)=\sum_{m=1}^n g_m(\ln(t/t_0))
\,.
\end{equation}
Because the leading order result $g_1(x)$ is of order ${\cal O}(1/\ln t)$
each higher order solution requires the inclusion of an additional line 
in Eq.~(\ref{eq:problemA4}). This adds new parameters like $\mu_6$ and 
$\kappa'$ in each step, whereas for the $A_3$-singularity, Eq.~(\ref{eq:gm}), 
additional parameters occur only in every second step of the expansion.

\subsection{\label{sec:discussion_one_A4}Discussion}

The results for the $A_4$-singularity shall be demonstrated for the kernel
\cite{Goetze1988b},
\begin{equation}\label{eq:F123} 
m(t) = v_1 \phi(t) +  v_2 \phi^2(t) + v_3 \phi^3(t) \,,
\end{equation}
substituted into the equation of motion~(\ref{eq:shorttime:Brown}) used with
$\tau=1$. The model has an $A_4$-singularity at $\mathbf{V}^*=(1,1,1)$ with
$f^*=0$ and coefficients $\mu_l,\,l\geqslant 4$ and $\kappa$ being unity.

\begin{figure}[htb]\includegraphics[width=0.9\columnwidth]{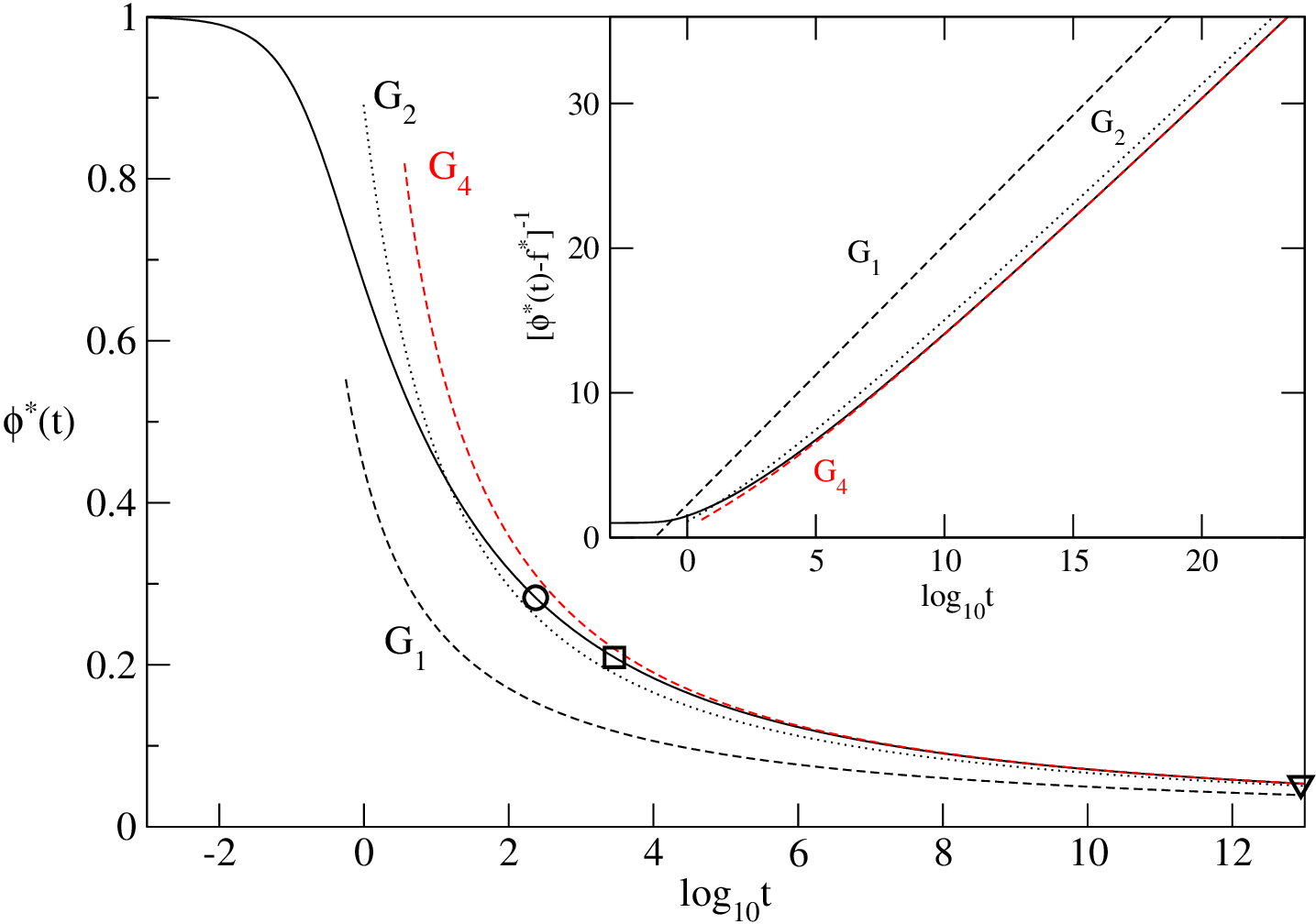}
\caption{\label{fig:F123}Critical decay $\phi^*(t)$ at the $A_4$-singularity 
of the model defined by Eqs.~(\ref{eq:shorttime:Brown}) and~(\ref{eq:F123}),
and the unit of time chosen such that $\tau=1$. The approximations by 
Eq.~(\ref{eq:A4sol}) with $t_0=0.055$ matched for $G_4$ are labeled 
accordingly. The square and the circle mark the time where the approximation 
by $G_4$ deviates from the solution by $5\%$ and $10\%$, respectively. The 
triangle refers to a $5\%$ deviation of $G_2$ from the solution. The inset 
displays the inverse of $[\phi^*(t)-f^*]$ and its respective approximations.
}
\end{figure}
Using up to four terms in the expansion~(\ref{eq:A4sol}), the time scale is 
fixed at $t_0=0.055$. Successive approximations to the numerical solution
are shown in Fig.~\ref{fig:F123}. Again, the leading approximation $G_1$ does 
not describe the solution.
The inset shows $[\phi^*(t)-f^*]^{-1}$, where a decay proportional to $1/\ln t$
would be seen as a straight line. $G_1$ yields such a straight line by 
definition; but it has the wrong slope compared to the solution. The latter 
exhibits a straight line for $t\gtrsim 10^7$. Including the leading 
correction in $G_2$ can account for the slope of the long-time solution.
Further terms in the asymptotic expansion enhance the accuracy of
the approximation. $G_4$ fulfills the $5\%$ criterion at $t=3\,10^3$, and is
in accord with the solution on the $10\%$ level for  $t>230$. $G_2$ intersects
$\phi^*(t)$ for shorter times but deviates first from the solution by $5\%$ at 
$t=9\cdot10^{12}$.

\section{\label{sec:critA3}Asymptotic expansion of the critical 
correlators at an $A_3$-singularity}

For the study of the general models, we go back to 
Eqs.~(\ref{eq:def_hatphi}--\ref{eq:def_Jqn}).
The solvability condition for Eq.~(\ref{eq:eom_hatphi}) reads
\numparts\label{eq:qsol_general}
\begin{equation}\label{eq:J_solvability}
\sum_q a^*_q J_q (z) = 0 \,\, ,
\end{equation}
and the general solution can be written as
\begin{equation}\label{eq:phisol}
\hat \phi_q (t) =   a_q \hat \phi (t) + \tilde \phi_q (t) \,\, .
\end{equation}
The splitting of $\hat \phi_q (t)$ in two terms is unique if one imposes 
the convention $\sum_q a^*_q \, \hat \phi_q (t) = \hat \phi (t)$. Then, the
part $\tilde\phi_q (t)$ can be expressed by means of the reduced resolvent 
$R_{q k}$ of $A_{q k}^{(1) c}$:
\begin{equation}\label{eq:a_perpendiclar}
{\cal S}[ \tilde \phi_q (t)] (z) = R_{q k} J_k (z)
\,\, .
\end{equation}\endnumparts
The matrix $R_{qk}$ can be evaluated from matrix $A_{q k}^{(1) c}$ and the 
vectors $a^*_k,\,a_k$ \cite{Gantmacher1974b}. Let us emphasize that 
Eqs.~(\ref{eq:J_solvability}--\textit{c}) together with the definitions in 
Eqs.~(\ref{eq:def_hatphi}) and~(\ref{eq:def_Jqn}) are an exact reformulation of
the equation of motion~(\ref{eq:MCT}) for states at glass-transition 
singularities. It is the aim of following 
calculations to express $\tilde{\phi}_q (t)$ recursively in terms of 
$\hat{\phi}(t)$ and to show that $\hat{\phi}(t)$ has the asymptotic expansion
discussed in Sec.~\ref{sec:crit_one_A3} for the one-component models. The 
starting point is the observation that $\tilde{\phi}_q (t)$ is small of higher 
order than $\hat{\phi}(t)$. This is obvious, since Eqs.(\ref{eq:def_Jqn})
and~(\ref{eq:a_perpendiclar}) imply $\tilde{\phi}_q(z) = {\cal O}(\hat{\phi}^2)
+ {\cal O}(\hat{\phi}\tilde{\phi}_q) + {\cal O}(\tilde{\phi}_q^2)$. Therefore,
one gets
\numparts
\begin{equation}\label{eq:JqzO2}
J_q(z)={\cal O}(\hat{\phi}^2)\,,
\end{equation}
\begin{equation}\label{eq:tildephiqO2}
\tilde{\phi}_q (t) = {\cal O}(\hat{\phi}^2)\,.
\end{equation}
\endnumparts
We assume that $\hat{\phi}$ can be expanded in terms of functions $g_m(x)$ as 
defined in Eqs.~(\ref{eq:logpoly_a},\textit{b}), and show the legitimacy of 
this Ansatz by the success of the following constructions.

\subsection{\label{sec:expansion_next_A3}Expansion up to next-to-leading 
order}

Substituting the splitting~(\ref{eq:phisol}) into the 
inhomogeneity $J^{(2)}_q(z)$ from Eq.~(\ref{eq:def_Jqn})
yields
\begin{equation}\label{eq:pihh2}
J_q(z) = A^{(2)c}_{qk_1k_2}a_{k_1}a_{k_2}{\cal S}[\hat{\phi}(t)^2](z)
- a_q^2{\cal S}[\hat{\phi}(t)]^2(z)+{\cal O}(\hat{\phi}^3)\,.
\end{equation}
The function $\psi_2(z)$ in Eq.~(\ref{eq:def_psi}) is of order 
${\cal O}(\hat{\phi}^3)$ because of Eq.~(\ref{eq:Tauber_cancellation}).
Therefore,
\begin{equation}\label{eq:phih2_reduced}
J_q(z) = (A^{(2)c}_{qk_1k_2}a_{k_1}a_{k_2}-a_q^2)\,
{\cal S}[\hat{\phi}(t)^2](z)+{\cal O}(\hat{\phi}^3)\,.
\end{equation}
Remembering Eq.~(\ref{eq:mu2}) and the condition $\mu_2=0$, one notices that
the solvability condition~(\ref{eq:J_solvability}) is fulfilled to
order ${\cal O}(\hat{\phi}^2)$. Hence, Eq.~(\ref{eq:a_perpendiclar}) yields 
\begin{equation}\label{eq:phit1}
\tilde{\phi}_q(t) = X_q\hat{\phi}^2(t)+ {\cal O}(\hat{\phi}^3)
\end{equation}
with the abbreviation \cite{Goetze2002}
\begin{equation}\label{eq:G2q_Xq}
X_q = R_{q k} \left[ A_{k k_1 k_2 }^{(2)c} a_{k_1} a_{k_2} -
a_{k}^2 \right]
\,\,.
\end{equation}
The first step in the derivation of $q$-dependent corrections results in
the extension of Eq.~(\ref{eq:phisol}):
\numparts\begin{equation}\label{eq:phiq1}
\hat{\phi}_q(t) = a_q\hat{\phi}(t) + X_q \hat{\phi}^2(t)
        + \tilde{\phi}'_q(t) \,,
\end{equation}
where
\begin{equation}\label{eq:phiq1_rest}
\tilde{\phi}'_q(t) = {\cal O}(\hat{\phi}^3)\,.
\end{equation}
\endnumparts

The next step is started by substituting the result~(\ref{eq:phiq1})
into  Eq.~(\ref{eq:def_Jqn}) for $J_q(z)$. Terms of order 
${\cal O}(\hat{\phi}^2)$ vanish altogether as demonstrated
above and only $a_q^2\psi_2(z)$ and additional terms of order ${\cal
O}(\hat{\phi}^3)$ are left from $J^{(2)}_q(z)$. Equation~(\ref{eq:def_psi}) 
is used to reduce products of ${\cal S}$-transforms to ${\cal S}$-transforms 
of products. The inhomogeneity assumes the form
\begin{eqnarray}\label{eq:phih3_reduced}\fl
J_q(z) = \,
{\cal S}[\hat{\phi}(t)^3](z)\,
        \left[A^{(3)c}_{qk_1k_2k_3}a_{k_1}a_{k_2}a_{k_3}
        +2(A^{(2)c}_{qk_1k_2}a_{k_1}X_{k_2}-a_q^2)
        -(a_q^3+2a_qX_q)\right]\nonumber\\
+\,a_q^2 \psi_2(z)
        +{\cal O}(\hat{\phi}^4)\,.
\end{eqnarray}
Let us introduce $\kappa=2\zeta$ and $\mu_3$ in agreement with 
Ref.~\cite{Goetze2002}:
\numparts
\begin{equation}\label{eq:zeta}
\zeta = \sum_q a^*_q \left[ a_q X_q +  a_q^3/2
\right] \,\, ,
\end{equation}
\begin{equation}\label{eq:mu3}
\mu_3 = 2 \zeta - \sum_q a^*_q \left[ A_{q k_1 k_2 k_3}^{(3)c}
a_{k_1} a_{k_2} a_{k_3} + 2  A_{q k_1 k_2}^{(2)c} a_{k_1}   
X_{k_2}
 \right] \,.
\end{equation}
\endnumparts
Then, the solvability condition~(\ref{eq:J_solvability}) reads
\begin{equation}\label{eq:solvability3}
0=\psi_2(z) - \mu_3 {\cal S}[\hat{\phi}(t)^3] + {\cal O}(\hat{\phi^4})\,.
\end{equation}
This equation was discussed 
in Sec.~\ref{sec:crit_one_A3}. The result is $\hat{\phi}(t) = g_2(x)+g_3(x)
+{\cal O}(1/x^4)$ with the functions $g_2(x)$ and $g_3(x)$ specified in
Eqs.~(\ref{eq:g2}) and ~(\ref{eq:g3}), respectively. From Eq.~(\ref{eq:phit1}) 
one infers, that $\tilde{\phi}_q(t) = {\cal O}(1/x^4)$. For the solution up to 
next-to-leading order, only the first term on the right-hand side of 
Eq.~(\ref{eq:phiq1}) matters. However, the discussion of the solvability 
condition including the $X_q\hat{\phi}^2(t)$-term was necessary in order to fix
the important number $\mu_3$, which enters Eq.~(\ref{eq:solvability3}) and 
thereby the cited formulas for $g_2(x)$ and $g_3(x)$.

\subsection{\label{sec:highcorrection}Higher-order expansions}

After substitution of Eq.~(\ref{eq:phiq1}) into Eq.~(\ref{eq:def_Jqn}) in order
to extend the expansion of $J_q(z)$, one can use Eq.~(\ref{eq:a_perpendiclar})
to determine $\tilde{\phi}'_q(t)$ up to errors of order 
${\cal O}(\hat{\phi}^4)$. There appears a new amplitude $Y_q$ as
\begin{equation}\label{eq:Yq}\fl\qquad
Y_q = R_{qk} \left\{\left[A_{kk_1k_2k_3}^{(3)c}a_{k_1}a_{k_2}a_{k_3} 
        - a_k^3\right] + 2 [A_{kk_1k_2}^{(2)c}a_{k_1}X_{k_2} - a_kX_k] 
        + \mu_3  a_k^2\right\}\,.
\end{equation}
To get the last term in the curly bracket, Eq.~(\ref{eq:solvability3}) was 
used to express the frequency dependence of $J_q(z)$ in 
Eq.~(\ref{eq:phih3_reduced}) solely by ${\cal S}[\hat{\phi}(t)^3](z)$.
After this second reduction step, one gets the extension of 
Eq.~(\ref{eq:phiq1}):
\numparts
\begin{equation}\label{eq:phiq2}
\hat{\phi}_q(t) = a_q\hat{\phi}(t) + X_q \hat{\phi}^2(t)
        + Y_q \hat{\phi}^3(t)   + \tilde{\phi}''_q(t) \,,
\end{equation}
where
\begin{equation}\label{eq:phiq2_rest}
 \tilde{\phi}''_q(t) = {\cal O}(\hat{\phi^4})\,.
\end{equation}
\endnumparts
Here, the contribution proportional to $Y_q$ has $g_2^3$ as the lowest-order 
term, and therefore it  is of higher order than $g_5$. However, the 
calculation of the amplitude $Y_q$ is a prerequisite to determine the 
parameter $\mu_4$, which will be needed below.

To continue, we substitute Eq.~(\ref{eq:phiq2}) into the 
solvability condition~(\ref{eq:J_solvability}). The same tricks as 
before are required to yield a definition of $\mu_4$ which is consistent 
with the equations for the one-component case. Before adding new terms 
from the expansion of $J_q(z)$ in Eq.~(\ref{eq:def_Jqn}), the 
remaining terms of order ${\cal O}(\hat{\phi}^5)$ in 
Eq.~(\ref{eq:phih3_reduced}) shall be collected from 
the lines with $n\leqslant 3$. A new parameter is 
introduced to shorten notation,
\begin{equation}\label{eq:kappadash}
\tilde{\kappa}=2 \sum_q a_q^* a_q X_q\,,
\end{equation}
and the contribution to $J_q(z)$
so far is $\kappa\psi_3(z)-\tilde{\kappa}{\cal S}[\hat{\phi}] 
\psi_2(z)$. Equation~(\ref{eq:solvability3}) can be used to eliminate 
$\psi_2(z)$. With the assistance of Eq.~(\ref{eq:def_psi}),
this contribution is reduced to $\kappa\psi_3(z) - \mu_3\tilde{\kappa} 
{\cal S}[\hat{\phi}^4]+{\cal O}(\hat{\phi}^5)$.
Next, the term from Eq.~(\ref{eq:def_Jqn}) for $n=4$ is added and the term
with $\tilde{\kappa}$ is absorbed in the definition of $\mu_4$. Then, the 
solvability condition reads
\begin{equation}
0=\kappa\psi_3(z) -\mu_4{\cal S}[\hat{\phi}^4] + {\cal O}(\hat{\phi^5})\,,
\end{equation}
where the definition for the remaining parameter $\mu_4$ is
\begin{eqnarray}\label{eq:mu4}\fl\qquad
\mu_4 = \sum_q a^*_q\{[a_q^4 - 
A_{qk_1k_2k_3k_4}^{(4)c}a_{k_1}a_{k_2}a_{k_3}a_{k_4}]
+3 [a_q^2 X_q - A_{qk_1k_2k_3}^{(3)c}a_{k_1}a_{k_2}X_{k_3}]\nonumber\\
+  [X_q^2 - A_{qk_1k_2}^{(2)c}X_{k_1}X_{k_2}]
+2 [a_q Y_q - A_{qk_1k_2}^{(2)c} a_{k_1} Y_{k_2}]
\} + \tilde{\kappa}\mu_3\,.\end{eqnarray}

After having defined all the necessary parameters, we see that the 
solution from Sec.~\ref{sec:highcorrection_one_A3} for $\hat{\phi}(t)$ is 
consistent with the solution of the $q$-dependent case as formulated in 
Eq.~(\ref{eq:phiq2}). Keeping only terms up to errors of order $(1/\ln t)^6$,
one arrives at the asymptotic formula for the critical correlator at an 
$A_3$-singularity,
\numparts
\begin{eqnarray}\label{eq:finalA3}\fl\qquad
\phi_q^\circ(t) =& f_q^\circ + h_q^\circ \{g_2(x) +g_3 (x)\nonumber \\
&\qquad +[g_4(x) + K^\circ_q g_2^2(x)] 
+ [g_5(x) +2 K^\circ_q g_2(x) g_3(x)] \}
\,,\end{eqnarray}
with\begin{equation}\label{eq:def_Kq}
h_q^\circ = (1-f^\circ_q) a_q
\,,\qquad K^\circ_q = X_q/a_q\,.
\end{equation}
\endnumparts
The first line of Eq.~(\ref{eq:finalA3}) expresses the factorization theorem:
$\phi_q^\circ(t) - f_q^\circ$ is a product of a first factor $h_q^\circ$, which
is independent of time, and a second factor $[g_2(x) +g_3 (x)]$, which is 
independent of the correlator index $q$. Factorization is first violated in 
order $1/\ln^4 t$, and only the terms with the amplitudes $K_q^\circ$ are
responsible for that. The expansion for $\phi_q^\circ(t)$ can be carried
out up to order $1/\ln^5 t$ if $\mu_4$ is known. The next order includes
$g_6(x)$ and requires knowledge of the additional parameter $\mu_5$.

\subsection{\label{sec:discussion_A3}Discussion}

As simple example for the demonstration of the preceding results, an $M=2$ 
model shall be considered. The MCT equations for Brownian dynamics read
for $q=1,\,2$
\numparts
\begin{eqnarray}\label{eq:BK_eom}\label{eq:BK_eom_int}
&&\tau_q \partial_t \phi_q(t) + \phi_q (t) +
\int_0^t m_q (t - t^\prime) \partial_{t^\prime} \phi_q
(t^\prime)dt^\prime = 0\,,\\
\label{eq:BK_eom_m1}
&&m_1(t) = v_1 \phi_1^2(t) + v_2 \phi_2^2(t)\,,\\
\label{eq:BK_eom_m2}
&&m_2(t) = v_3 \phi_1(t)\phi_2(t)\,\,.
\end{eqnarray}
\endnumparts
This is a schematic model for a symmetric molten salt \cite{Bosse1987b}. The 
model has three control parameters, $\mathbf{V}=(v_1, v_2, v_3)$. The 
glass-transition singularities in this system can be evaluated analytically.
There is an $A_4$-singularity at $v^*_3\approx 24.78$, and $A_3$-singularities 
occur for $v_3>v_3^*$. To allow for a comparison with previous work 
\cite{Goetze2002}, we set $\tau_1=\tau_2=1$ and choose the $A_3$-singularity 
for $v_3^\circ=45$.

\begin{figure}[htb]\includegraphics[width=0.9\columnwidth]{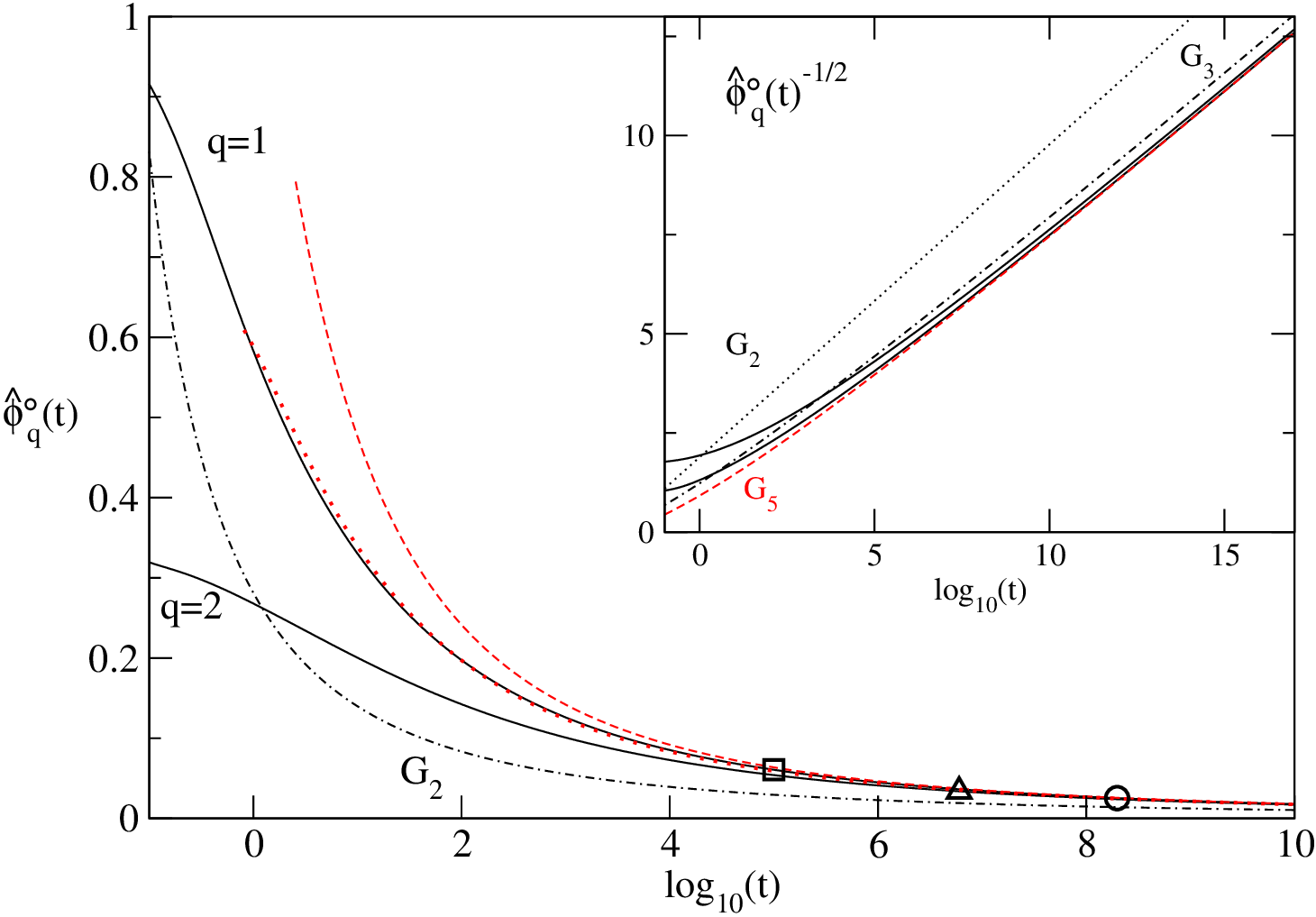}
\caption{\label{fig:BK45}Critical decay at the $A_3$-singularity in the 
two-component model defined by Eqs.~(\ref{eq:BK_eom}--\textit{c}) for 
$v_3^\circ=45$ and $\tau_1=\tau_2=1$. The rescaled solutions 
$\hat{\phi}^\circ_q(t) = [\phi^\circ_q(t)-f^\circ_q]/h_q^\circ$, $q=1,2$, are 
shown for as full lines.  
The asymptotic approximation~(\ref{eq:finalA3}) is shown dashed for
$q=1$ and dotted for $q=2$. The points where the approximation deviates by
$5\%$ from the solution for $q=1,2$, and the point where the solutions 
differ by $5\%$ from each other are marked by a square, a triangle and a 
circle, respectively. The chain curve with label $G_2$ shows the leading 
contribution from Eq.~(\ref{eq:finalA3}).
The inset shows as full lines the rectification,
$\hat{\phi}^\circ_q(t)^{-1/2}$ for $q=1$ (lower full line) and $q=2$ 
(upper full line). The $q$-independent part $G_5$ of the approximation in
Eq.~(\ref{eq:finalA3}) is given by the dashed line. The dotted and the
chain line show the leading and next-to-leading order approximations $G_2$
and $G_3$, Eq.~(\ref{eq:finalA3}). The time scale $t_0$ is $4.07\cdot
10^{-3}$.
}
\end{figure}
Let us use the rescaled correlators $\hat{\phi}_q^\circ(t)=[\phi^\circ_q(t)
-f^\circ_q]/h_q^\circ$ for the following considerations. The result in 
Eq.~(\ref{eq:finalA3}) assumes the form $\hat{\phi}_q^\circ(t)=G_5(x)
+K_q^\circ\tilde{G}_{5}(x)$, with $G_5(x)$ from Eq.~(\ref{eq:sol_n}) and 
$\tilde{G}_{5}(x)=g_2^2(x)+2g_2(x)g_3(x)$. Since $\tilde{G}_{5}(x)$ is of 
higher order than $G_5(x)$, Eq.~(\ref{eq:g2}), 
correlators for different $q$ approach each other for sufficiently large time 
as is demonstrated in Fig.~\ref{fig:BK45}. The time $t\approx2\cdot 10^8$, 
where $\hat{\phi}_2^\circ$ deviates by $5\%$ from $\hat{\phi}_1^\circ$, is 
marked by a circle. The amplitude $K_q^\circ$ introduces the $q$-dependent
corrections which are smaller for $q=1$ than for $q=2$. To evaluate $G_5(x)$
and $\tilde{G}_{5}(x)$, we determined the following parameters, 
$\mu_3=0.772$, $\kappa=0.888$, and $\mu_4=1.38$. Notice, that $\mu_3$ is more
than twice as for the model studied in 
Fig.~\ref{fig:F13}. Since the coefficients $c_{m,l}$ in Eq.~(\ref{eq:gm}) 
contain powers of $\mu_3$ in the denominator, corrections 
are smaller if $\mu_3$ is larger, cf. Eqs.~(\ref{eq:c40})--(\ref{eq:c42}) 
and~(\ref{eq:c50})--(\ref{eq:c53}). Because of the smaller corrections,
the time scale can be matched with $G_5(x)$
between $t=10^{20}$ and $10^{25}$ which is significantly earlier than for 
the model studied in Fig.~\ref{fig:F13}. We get $t_0=4.07\cdot 10^{-3}$.

The asymptotic approximation~(\ref{eq:finalA3}) is shown as a dashed
line for $q=1$ in Fig.~\ref{fig:BK45}, it deviates by more than
$5\%$ from the solution if $t\lesssim 10^5$ ($\square$). The approximation
for $q=2$ (dotted) deviates by more than $5\%$ for $t\lesssim 6\cdot 10^6$
($\triangle$). This difference in the range of validity can be understood
qualitatively by considering the $q$-dependent corrections of higher order
in Eq.~(\ref{eq:phiq2}), $K_q[g_3^2(x)+2g_2(x)g_4(x)]+Y_qg_2^3(x)/a_q$
with $Y_q$ from Eq.~(\ref{eq:Yq}). Both $K_q$ and $Y_q/a_q$ are
smaller for the first correlator, $Y_1/a_1=-0.1928$ and $Y_2/a_2=5.761$,
and introduce less deviations from the $q$-independent part $G_6(x)$ of
the approximation in higher order.

The $q$-independent function $G_5(x)$ would lie on top of the dashed line
in Fig.~\ref{fig:BK45} and is therefore shown only in the inset
which also displays the critical correlators and the $q$-independent
functions $G_2(x)$ and $G_3(x)$, Eq.~(\ref{eq:sol_n}). Plotting
$\hat{\phi}^\circ_q(t)^{-1/2}$ we can identify $1/\ln^2 t$-behavior as
straight line. The critical correlators exhibit a straight line starting
from $t\approx 10^9$. The leading approximation $G_2(x)$ is a straight
line as well but has a slope slightly larger than the solution. The
first correction $G_3(x)$ resembles the slope of the solution but is
offset from the solution by a shift of the time scale. This was observed
before in Fig.~\ref{fig:F13}. Since $G_5(x)+K^\circ_q\tilde{G}_5(x)$ was used
to match the time scale $t_0$ and as $\tilde{G}_5(x)$ decays faster than the
$q$-independent part, $G_5(x)$ coincides with the solution for larger
times.

As a second example, the
asymptotic laws shall be considered for the square-well system (SWS). This is 
the microscopic model for a colloid explained in Sec.~\ref{sec:intro}. The 
microscopic version of MCT for colloids is used with the wave-vector moduli
discretized to a set of $M=500$ values. The structure factors that define the 
mode-coupling functional ${\cal F}_q$ in Eq.~(\ref{eq:def_m}) are calculated
in the mean-spherical approximation. We shall consider the same 
$A_3$-singularity for $\delta^\circ=0.03$ as considered in previous studies
\cite{Sperl2003a,Sperl2004}. The reader is referred to these papers for further
details and for an extensive discussion of the relaxation near the specified
$A_3$-singularity. For the 
evaluation of the approximation~(\ref{eq:finalA3}), we need the correction 
amplitudes $K^\circ_q$ which are shown in Fig.~8 of Ref.~\cite{Sperl2003a} and
the parameters characteristic for the $A_3$-singularity under discussion,
\begin{equation}\label{eq:A3mu4}
\mu_3 = 0.109,\,\quad\kappa=0.314\,,\quad
\mu_4 = 0.204\,.
\end{equation}
The asymptotic approximation reads
\begin{eqnarray}\label{eq:SWSA3crit}\fl\quad
\hat{\phi}^\circ_q(t)
=\, 60.4/x^2 +264.7\ln x/x^3\nonumber\\
        + [3374.9-580.2\ln x+870.4\ln^2x]/x^4\nonumber\\
        + [-11745.7-27952.1\ln x-4452.2\ln^2x+2544.1\ln^3x]/x^5\nonumber\\
        +K^\circ_q\left\{3643.9/x^4+31953.7\ln x/x^5\right\}
+{\cal O}(x^{-6})\,.
\end{eqnarray}
The first line represents $g_2(x)$ and 
$g_3(x)$, Eqs.~(\ref{eq:g2}) and~(\ref{eq:g3}). The second and 
third line exhibit the contributions up to $g_4(x)$ and $g_5(x)$, 
Eqs.~(\ref{eq:gm}-\ref{eq:c53}), which are independent of the wave vector. 
The $q$-dependent correction terms appear with the prefactor $K^\circ_q$ in the
curly brackets; they are positive for $t/t_0 > 2.5$ and monotonically 
decreasing for $t/t_0> 3.1$.

\begin{figure}[htb]\includegraphics[width=0.9\columnwidth]{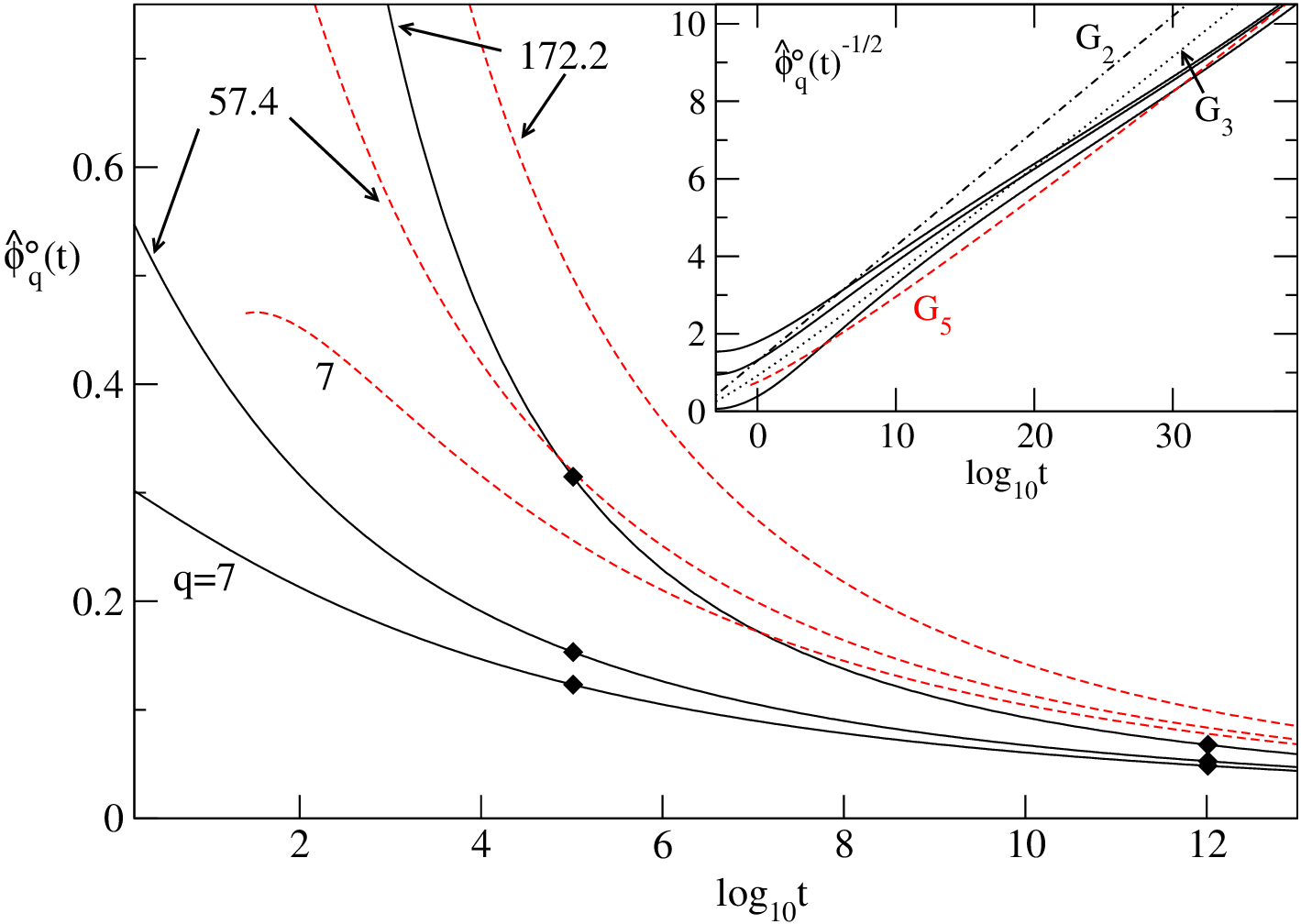}
\caption{\label{fig:SWSA3}Critical decay at the $A_3$-singularity of the 
square-well system (SWS) for the relative attraction-shell width $\delta^\circ
=0.03$. Full lines show the rescaled correlation functions 
$\hat{\phi}^\circ_q(t)=[\phi^\circ_q(t)-f_q^\circ]/h_q^\circ$ at 
$\mathbf{V}^\circ$, for the wave-vector values $qd=7,\,57.4,$ and $172.2$ as 
indicated. The unit of time is chosen so that $1/D_0=160$, with $D_0$ denoting
the single particle diffusivity \cite{Sperl2003a,Sperl2004}.
The dashed lines exhibit the asymptotic approximation of
Eq.~(\ref{eq:SWSA3crit}) with a time scale $t_0=4\cdot 10^{-5}$
matched in the interval $t=10^{40}$\dots$10^{45}$. For $qd=7.0$, $57.4$,
and $qd=172.2$, the correction amplitudes are $K^\circ_q=-1.704$, 
$-0.00224$, and $4.814$, respectively. The filled diamonds for $t=10^5$ 
and $t=10^{12}$
mark the values for $\hat{\phi}^\circ_q(t)$ for the three $q$-values. The
inset shows $\hat{\phi}^\circ_q(t)^{-1/2}$ for the $q$-values above from 
top to bottom and the $q$-independent approximations defined in 
Eq.~(\ref{eq:sol_n}) in
the same representation, $G_2(t)^{-1/2}$, $G_3(t)^{-1/2}$ and 
$G_5(t)^{-1/2}$, respectively. 
}
\end{figure}
Figure~\ref{fig:SWSA3} shows the rescaled functions $\hat{\phi}^\circ_q(t)$ for
three representative wave numbers. At the peak of the structure factor, $qd=7$,
the amplitude is negative, for $qd=57.4$ the correction amplitude is close to 
zero, and for the wave vector $qd=172.2$ the amplitude is positive. The 
functions (full lines) deviate strongly from each other in the window of time 
presented, demonstrating severe violation of the  factorization property. If 
the deviations among the
correlation functions for different wave vectors cannot be assigned to the
$q$-dependent corrections in Eq.~(\ref{eq:SWSA3crit}) within an accessible 
window in time, we cannot expect that Eq.~(\ref{eq:SWSA3crit}) will be 
sufficient to describe the critical decay.
Suppose, the critical correlators for different
wave vectors are approximated by Eq.~(\ref{eq:SWSA3crit}). Then, for
arbitrarily chosen wave vectors $q_1$ and $q_2$, the difference
$\hat{\Delta}[q_1,q_2](t)=\hat{\phi}^\circ_{q_1}(t)-\hat{\phi}^\circ_{q_2}(t)$
is given in leading order by the difference in the correction amplitudes,
$K^\circ_{q_1}-K^\circ_{q_2}$, and the terms in the curly brackets in
Eq.~(\ref{eq:SWSA3crit}). From Fig.~\ref{fig:SWSA3} we infer that 
$\hat{\Delta}[q_1,q_2](t)$ is not yet close to zero to neglect the terms in the
curly brackets. The values of $\hat{\phi}^\circ_q(t)$ for the three chosen 
$q$-values are marked by diamonds in Fig.~\ref{fig:SWSA3} for $t=10^5$ and
$10^{12}$. We get $\hat{\Delta}[7,57.4](10^5)=-0.030$ and 
$\hat{\Delta}[172.2,57.4](10^5)=0.161$. These differences are large 
but they correctly reflect the ordering in the values for $K^\circ_q$ which 
increase with $q$. From that we conclude that the treatment of the 
$q$-dependence in Eq.~(\ref{eq:SWSA3crit}) is qualitatively 
correct.

If the time dependence of $\hat{\Delta}[q_1,q_2](t)$ 
were given exclusively by the terms in curly brackets in 
Eq.~(\ref{eq:SWSA3crit}), then the differences among the $K_q^\circ$ would 
explain the amplitudes of the decay in $\hat{\Delta}[q_1,q_2](t)$. To quantify
deviations from that case we introduce the ratio $\nu[q_1,q_2,q_3](t)=
\hat{\Delta}[q_1,q_2](t)/\hat{\Delta}[q_2,q_3](t)$. For $t\rightarrow\infty$ 
this ratio is 
$\nu_\infty=(K_{q_1}^\circ-K_{q_2}^\circ)/(K_{q_2}^\circ-K_{q_3}^\circ)$.
Deviations from $\nu_\infty$ indicate that higher-order
$q$-dependent corrections are present in addition to the terms in
Eq.~(\ref{eq:SWSA3crit}). For the $q$-values used in
Fig.~\ref{fig:SWSA3} we get $\nu_\infty= 
(K^\circ_{7}-K^\circ_{57.4})/(K^\circ_{57.4}-K^\circ_{172.2})=0.354$.  
Since $K_{57.4}^\circ\approx 0$, this ratio is almost equivalent to 
$-K^\circ_7/K^\circ_{172.2}$. 
The ratio at time $t=10^5$ is $\nu[7,57.4,172.2](10^5)=0.187$ and 
therefore deviates by $90\%$ from $\nu_\infty$. Hence, we cannot expect 
Eq.~(\ref{eq:SWSA3crit}) to describe the critical decay in 
Fig.~\ref{fig:SWSA3} at that time.
At $t=10^{12}$, the ratio has decayed to $\nu[7,57.4,172.2](10^{12})=0.280$
which which deviates from $\nu_\infty$ by $20\%$. Here, the $q$-dependent
corrections are also in reasonable quantitative agreement with the
approximation in Eq.~(\ref{eq:SWSA3crit}).
To determine $t_0$, we use extremely large times. The inset of
Fig.~\ref{fig:SWSA3} displays the rescaled correlators as
$\hat{\phi}^\circ_q(t)^{-1/2}$. In this representation, the leading 
term
$g_2(x)$ in Eq.~(\ref{fig:SWSA3}) yields a straight line. We see
that for large times the correlators for different $q$ indeed come closer 
together and the ratio at $t=10^{40}$ is 
$\nu[7,57.4,172.2](10^{40})=0.341$, which deviates by $4\%$ from $\nu_\infty$.
For the determination of $t_0$ we use
Eq.~(\ref{fig:SWSA3}) for $q=7,\,57.4\,,$ and $172.2$ and match
the asymptotic approximation to the numerical solutions in the interval
from $t=10^{40}$ to $t=10^{45}$. This results in a value $t_0=4\cdot
10^{-5}$. 
For times larger than $t\approx 10^{50}$ the numerical solution does no longer 
follow the approximation. In that region inaccuracies in the control-parameter
values lead to deviations from the asymptotic behavior. These inaccuracies 
prevent us also from fixing more than just one digit of $t_0$. 
The dashed line in the inset labeled
$G_5$ shows the result for neglecting the last line of
Eq.~(\ref{eq:SWSA3crit}). This also describes
the correlator for $q=57.4$ where $K_q$ is close to zero. Taking into
account only the first line of Eq.~(\ref{eq:SWSA3crit})
yields the dotted curve labeled $G_3$. This curve is
clearly inferior to $G_5$, but it captures the slope of the solution still
better than $G_2$. 

In the large panel of Fig.~\ref{fig:SWSA3}, one can compare the
critical correlators with the approximation by Eq.~(\ref{eq:SWSA3crit}). For
times of interest for experimental studies, the description is reasonable 
qualitatively. Especially the leading $q$-dependent corrections
describe the variations seen in the correlators down to relatively short times.
The accuracy of the approximation that was demonstrated for the schematic
models in Figs.~\ref{fig:BK45} and~\ref{fig:F13} is far
better than seen in Fig.~\ref{fig:SWSA3} for the SWS. This
difference is mainly due to different values of the parameter $\mu_3$ that
characterizes the various $A_3$-singularities. For the two-component model
we had $\mu_3=0.77$ and for the one-component model there was $\mu_3=1/3$.
The small value $\mu_3=0.109$ for the SWS implies slow convergence of the 
asymptotic expansion. Therefore, a quantitative description by 
Eq.~(\ref{eq:SWSA3crit}) is possible only for times exceeding considerably the
ones shown in Fig.~\ref{fig:SWSA3}.

\section{\label{sec:critA4}Asymptotic expansion of the critical 
correlators at an $A_4$-singularity}

\subsection{\label{sec:expansion_next_A4}Expansion up to next-to-leading 
order}

The calculation of the critical correlator at the $A_4$-singularity is so 
involved, that we restrict ourselves to the leading and next-to-leading order
result. The Eqs.~(\ref{eq:JqzO2}--\ref{eq:mu4}) remain valid, and 
Eqs.~(\ref{eq:Yq}) and ~(\ref{eq:mu4}) simplify because $\mu_3=0$. The 
difficulty comes about because $\mu_5$, which enters Eq.~(\ref{eq:A4g2}), has
to be determined. This requires the extension of Eq.~(\ref{eq:phiq2}), and 
thereby there appears a further amplitude.
The additional amplitude $Z_q$ is obtained by also including terms with $n=4$ 
from Eq.~(\ref{eq:def_Jqn}). Applying the same manipulations as above, one 
arrives at $\tilde{\phi}''_q(t)= Z_q\hat{\phi}^4+{\cal O}(\hat{\phi}^5)$ with 
the amplitude
\begin{eqnarray}\label{eq:Zq}\fl
Z_q = R_{qk}\{
[A_{kk_1k_2k_3k_4}^{(4)c}a_{k_1}a_{k_2}a_{k_3}a_{k_4}-a_k^4]
+3 [A_{kk_1k_2k_3}^{(3)c}a_{k_1}a_{k_2}X_{k_3}-a_k^2 X_k]\nonumber\\
+  [A_{kk_1k_2}^{(2)c}X_{k_1}X_{k_2}-X_k^2]
+2 [A_{kk_1k_2}^{(2)c} a_{k_1} Y_{k_2}-a_k Y_k]
+ \mu_4 a_k^2
\} \,.
\end{eqnarray}
Introducing the third $q$-dependent correction, the solution assumes the form
\begin{equation}\label{eq:phih5}
\hat{\phi}_q(t) = a_q \hat{\phi}(t) + X_q\hat{\phi}^2(t)
 + Y_q\hat{\phi}^3(t) + Z_q\hat{\phi}^4(t)+{\cal O}(\hat{\phi}^5)\,.
\end{equation}
Collecting all terms of order ${\cal O}(\hat{\phi}^4)$ after including 
also the line $n=5$ from Eq.~(\ref{eq:def_Jqn}), one gets from the solvability
condition~(\ref{eq:J_solvability}):
\begin{eqnarray}\label{eq:mu5_def}
\mu_5 =& \sum_q\,a^*_q\{
[a_k^5-A_{kk_1k_2k_3k_4k_5}^{(5)c}a_{k_1}a_{k_2}a_{k_3}a_{k_4}a_{k_5}]
\nonumber\\&
+4[a_k^3X_k - A_{kk_1k_2k_3k_4}^{(4)c}a_{k_1}a_{k_2}a_{k_3}X_{k_4}]
\nonumber\\&
+3[a_k X_k^2 +a_k^2Y_k- A_{kk_1k_2k_3}^{(3)c} 
        (a_{k_1}X_{k_2}X_{k_3} + a_{k_1}a_{k_2}Y_{k_3})]
\nonumber\\&
+ 2 [X_kY_k+a_kZ_k - A_{kk_1k_2}^{(2)c}(X_{k_1}Y_{k_2}+a_{k_1}Z_{k_2})]
\} + \tilde{\kappa}\mu_4\,.
\end{eqnarray}

Summarizing, the asymptotic solution for the critical decay at an 
$A_4$-singularity in next-to-leading order reads
\begin{equation}\label{eq:finalA4}
\phi_q^*(t) = f_q^* + h_q^* \left\{g_1(x)+[g_2(x)+K_q^* g_1^2(x)]\right\}\,.
\end{equation}
Here, in analogy to Eq.~(\ref{eq:def_Kq}), the critical amplitude is 
$h_q^*=(1-f^*_q)a_q$ and the correction amplitude is given by $K_q^*=X_q/a_q$.
The factorization theorem is obeyed by the leading-order term only. 
Contrary to what was found in Eq.~(\ref{eq:finalA3}) for the behavior at the 
$A_3$-singularity, already the leading correction term $g_2$ is modified by the
$q$-dependent term $K^*_q g_1^2(x)$ of the same order. The higher-order 
contributions enter the curly brackets in Eq.~(\ref{eq:finalA4}) as 
$g_3(x)+2g_1(x)g_2(x)X_q/a_q+g_1^3(x)Y_q/a_q$ and 
$g_4(x)+g_2^2(x)X_q/a_q+2g_1(x)g_3(x)X_q/a_q+3g_1^2(x)g_2(x)Y_q/a_q
+g_1^4(x)Z_q/a_q$. However, $g_3(x)$ requires the evaluation of the 
parameters $\mu_6$ and $\kappa'$, $g_4(x)$ needs $\mu_7$ and $\kappa''$.

\subsection{\label{sec:discussion_A4}Discussion}

\begin{figure}[htb]\includegraphics[width=0.9\columnwidth]{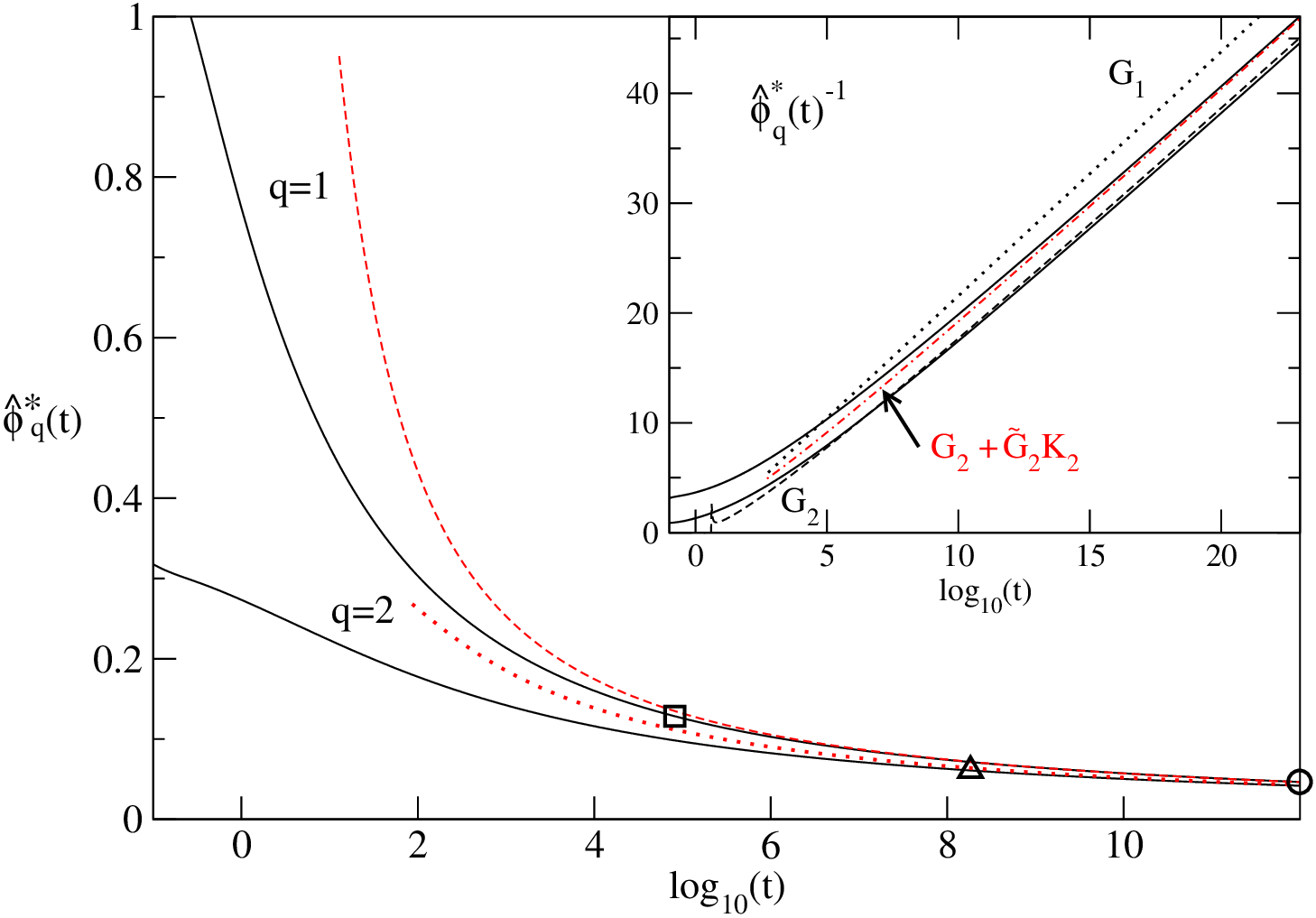}
\caption{\label{fig:BKA4}Rescaled critical decay $\hat{\phi}_q^*=[\phi^*_q(t)
-f_q^*]/h_q^*$ at the $A_4$-singularity in the two-component model defined 
in Eqs.~(\ref{eq:BK_eom}--\textit{c}) (full lines). 
The asymptotic approximations,
Eq.~(\ref{eq:finalA4}), for $q=1, 2$, are represented by the dashed
and dotted curve, respectively. For $q=1$ ($\square$) and $q=2$
($\triangle$), the points are marked where the solution and the
approximation deviate by $5\%$. An additional point is indicated where the
solution for $q=2$ differs from the one for $q=1$ by $10\%$ ($\bigcirc$).
The inset displays the rectified representation of the solutions for $q=1$
(lower full line) and $q=2$ (upper full line) together with the
$q$-independent parts of the approximations, $G_1$ and $G_2$, cf.
Eq.~(\ref{eq:A4sol}), and $G_2+K_2\tilde{G}_2$ (see text). The time 
scale $t_0=2.0$ was matched for $t=10^{20}\dots 10^{25}$.
}
\end{figure}
Figure~\ref{fig:BKA4} shows the critical decay at the $A_4$-singularity of the 
two-component model defined in Eqs.~(\ref{eq:BK_eom}--\textit{c}). The 
parameters for the evaluation of $g_1(x)$ and $g_2(x)$ are $\mu_4=1.53$, 
$\mu_5=0.962$, and $\kappa=0.386$. We use again the rescaled correlator
$\hat{\phi}^*_q(t)=[\phi^*_q(t)-f_q^*]/h_q^*$ and check first the validity
of the factorization in Eq.~(\ref{eq:finalA4}) in the form $\hat{\phi}^*_q(t)=
G_2(x)+K_q\tilde{G}_{2}(x)$ where $G_2(x)=g_1(x)+g_2(x)$ and $\tilde{G}_{2}(x)=
g_1^2(x)$. The time, where the solutions for $q=1,\,2$, differ by $5\%$ is only
reached at $t\approx 10^{23}$. The circle marks the point where the deviation 
is still $10\%$ at $t=10^{12}$. We can then use the 
approximation~(\ref{eq:finalA4}) 
to fix the time scale to $t_0=2.0$ which then yields the dashed and dotted 
curves for $q=1,\,2$, accordingly. For $q=1$ this approximation deviates 
by $5\%$ from the solution at $t\approx 8.2\cdot 10^4$ ($\square$). For 
$q=2$ we find $t\approx 1.8\cdot 10^8$ ($\triangle$). This is 
plausible when appealing to the $q$-dependent higher-order correction in
Eq.~(\ref{eq:phih5}), which incorporates in addition to drastically 
different values for $K_q$ also the values $Y_1/a_1=-0.579$ and 
$Y_2/a_2=3.76$. A rectified representation of the critical decay and the 
approximation in the inset shows again the 
leading-order $G_1(x)$ (dotted) as a straight line of different slope than 
the solution (full lines) and the second correction $G_2(x)$ (dashed). In 
this plot, the critical correlators for different $q$ are still 
significantly different in the entire window. But 
Eq.~(\ref{eq:finalA4}) can account for that difference as is shown by 
the good agreement of the curve labeled $G_2+\tilde{G}_2K_2$. The latter
describes the second correlator where the deviations due to 
the correction amplitudes are largest.

\begin{figure}[htb]\includegraphics[width=0.9\columnwidth]{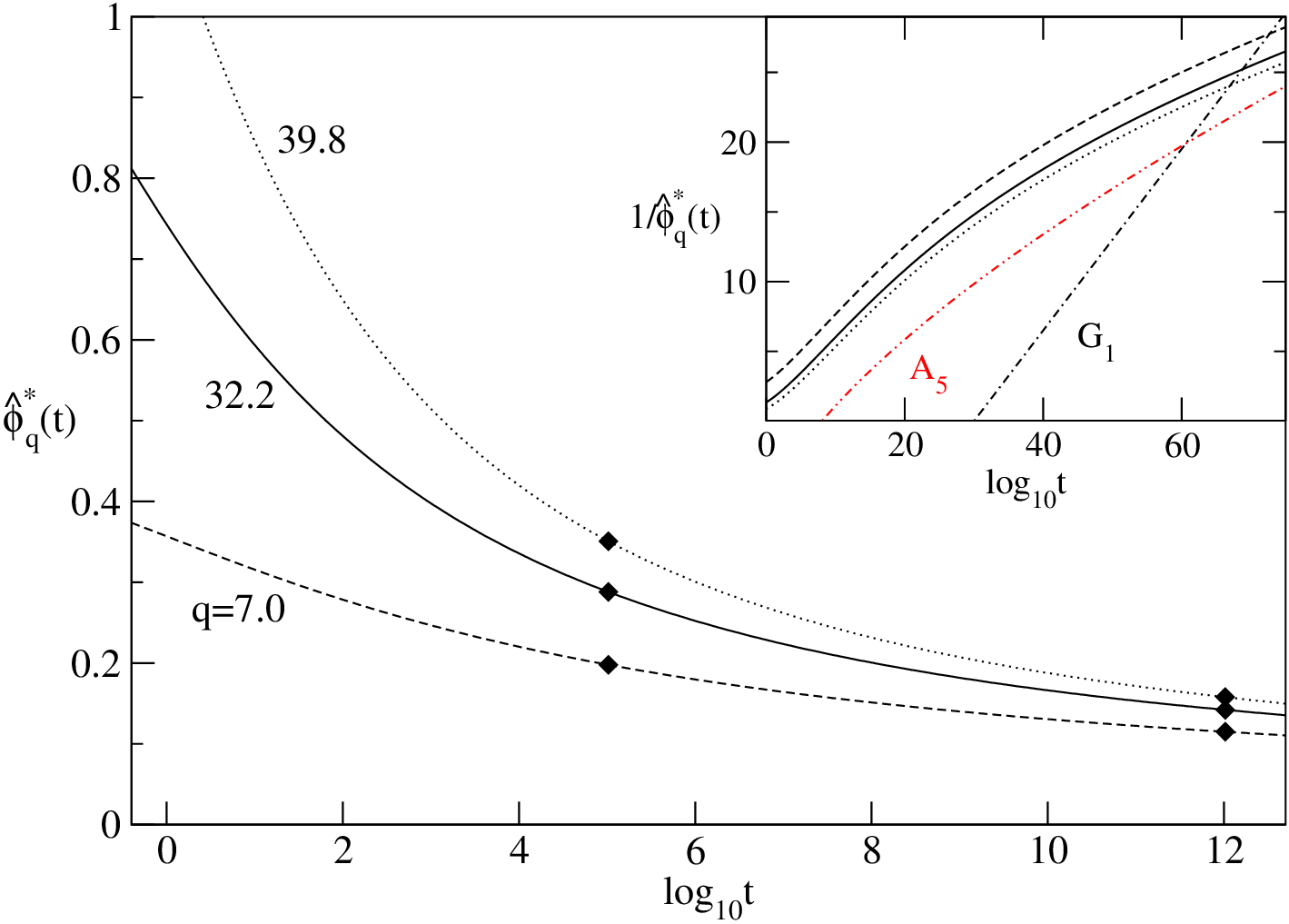}
\caption{\label{fig:SWSA4}Critical decay at the $A_4$-singularity of the SWS 
for $qd=7.0$ (dashed), $32.2$ (full line), $39.8$ (dotted). The correction
amplitudes are $K_q=-1.81$, $-0.04$, and $0.77$, respectively.
The filled diamonds mark the values at $t=10^5$ and $t=10^{12}$ where the 
ratios $\nu(t)$ are $1.44$ and $1.72$, respectively (see text).
The inset replots the curves from the full panel in the same linestyle and 
shows the first term of Eq.~(\ref{eq:SWSA4crit}) labeled $G_1$ and 
the law $\ln(t/\tau)^{-2/3}$ labeled $A_5$, both with an arbitrary time 
scale.
}
\end{figure}
We now turn to the $A_4$-singularity of the SWS. 
For the application of Eq.~(\ref{eq:finalA4}) we need the parameters 
characterizing the $A_4$-singularity,
\begin{equation}\label{eq:A4mu45}
\mu_4 = 0.131\,,\quad
\kappa = 0.243\,,\quad
\mu_5 = 1.21\,.
\end{equation}
The rather small value of $\mu_4$ generates particularly large coefficients in 
the expansion of the critical decay in Eq.~(\ref{eq:A4:gm}) where $\mu_4$ 
appears in the denominators. This feature is quite the same as mentioned above
for the $A_3$-singularities. The asymptotic approximation in 
Eq.~(\ref{eq:finalA4}) yields for the critical decay of the rescaled 
correlators:
\begin{equation}\label{eq:SWSA4crit}\fl\qquad
\hat{\phi}_q^*(t)
= \;3.54/x-50.7\ln x/x^2+12.5\, K_q^*/x^2
+{\cal O}(x^{-3})\,.
\end{equation}
We chose again values for $q$ where $K^*_q$ is negative, almost zero and 
positive. Figure~\ref{fig:SWSA4} demonstrates that the factorization is 
strongly violated.
Comparing the solutions $\hat{\phi}_q^*(t)$ for $t=10^5$ we find a ratio 
defined as in the previous section of $\nu[7,32.2,39.8](10^5)=1.439$ which 
is more than $30\%$ off the ratio for the correction 
amplitudes $\nu_\infty=2.185$. At $t=10^{12}$ we find  
$\nu[7,32.2,39.8](10^{12})=1.723$ which achieves $20\%$ accuracy.
So the critical decay at the $A_4$-singularity shown in   
Fig.~\ref{fig:SWSA4} is in qualitative accord with 
Eq.~(\ref{eq:SWSA4crit}) with respect to the variation in $q$.
However, due to the small value of $\mu_4$, the differences among the 
correlators for different $q$ do not decay fast enough to allow for a 
consistent determination of $t_0$ for the maximum value in time that could 
be reached. Numerically we find $\nu[7,32.2,39.8](10^{128})=2.076$ which is 
still $5\%$ off from $\nu_\infty$, and $\hat{\phi}^*_q(t)$ itself deviates 
from zero by $5\%$. This illustrates drastically the enormous stretching at the
$A_4$-singularity.

The inset of Fig.~\ref{fig:SWSA4} demonstrates that the critical decay 
$\hat{\phi}^*_q(t)$ is qualitatively different from the leading order 
$1/\ln t$-law for $t\leqslant 10^{60}$. For the $A_3$-singularity
in Fig.~\ref{fig:SWSA3} it was still possible to argue that curve
$G_2$ is in accord with the decay qualitatively at least for large times
and to attribute deviations for shorter times to the proximity of the
$A_4$-singularity. Figure~\ref{fig:SWSA4} does not allow for such an
interpretation. The curves $1/\hat{\phi}_q^*(t)$ have a slope smaller than
$1/G_1$ over the complete window in time and imply a slower decay than 
given by the leading order in Eq.~(\ref{eq:SWSA4crit}). If $\mu_4$ was zero,
the singularity would be of type $A_5$. The leading order critical decay at
such butterfly singularity is $\ln(t)^{-2/3}$. This law is added in the inset 
as chain line labeled $A_5$. Indeed, it explains the data qualitatively. Hence,
the shortcomings of the asymptotic expansion at the $A_4$-singularity in the
SWS result from the small value of $\mu_4$. 

To check if the value for $\mu_4$ is exceptionally small for the SWS, the 
calculation was repeated for the hard-core Yukawa system as introduced in 
Ref.~\cite{Goetze2003b}. We find the even smaller value $\mu_4=0.080$. 
Therefore, the small value of $\mu_4$ seems to be typical for systems with 
short-ranged attraction.

\section{\label{sec:summary}Summary}

The asymptotic expansion for large times of the critical decay of correlation
functions at higher-order glass-transition singularities has been elaborated. 
These decays can be considered as the analogue of the $t^{-a}$-law expansion 
for the correlators at the liquid-glass transition. The latter as well as the
higher-order singularities are obtained as bifurcations of type 
$A_l,\,l\geqslant 2$. The $A_l$-singularity and especially the critical decay 
law at the singularity is characterized by a number $\mu_l$. For the 
$A_2$-singularity of the liquid-glass transition, this characteristic number
determines the so-called exponent parameter $\lambda=1-\mu_2$, which specifies 
the critical exponent $a$ via $\lambda=\Gamma(1-a)^2/\Gamma(1-2a)$. For 
$\mu_2=0$ or $\lambda=1$, one gets $a=0$ and the asymptotic expansion in terms 
of powers $t^{-a}$ becomes invalid. A higher-order singularity $A_n$ is 
encountered, defined by $\mu_n>0$ while $\mu_l=0$ for $l<n$. 

For an $A_3$-singularity, the critical decay is given by an expansion in 
inverse powers of the logarithm of the time, starting with $1/\ln^2t$. The 
convergence of the asymptotic expansion is the better the larger is $\mu_3$. 
The result for the general models in Eqs.~(\ref{eq:finalA3}) and~(\ref{eq:gm})
adds probing-variable dependent correction terms to the one-component result.
These can be expressed by terms from the one-component solution and correction
amplitudes. The leading correction amplitude $K_q$ is the same function of the 
MCT-coupling constants as found earlier for the logarithmic decay-law 
expansions 
\cite{Goetze2002}. Since the vertex is a smooth function of the control 
parameters, these correction amplitudes are smooth functions as well. 
Therefore, also for the general case, the range of validity for the asymptotic 
expansion is determined by the characteristic parameter
$\mu_3$. If $\mu_3$ is small, the quality of the fit by the asymptotic
expansion is less satisfactory than for larger $\mu_3$. Generically, larger
$\mu_3$ can be obtained by extending the corresponding glass-glass-transition
line deeper into the glassy region and hence having the $A_3$-singularity 
further separated from the liquid regime. Thus, the
dynamics influenced by an $A_3$-singularity seen in the liquid regime is
either connected to a rather small $\mu_3$, or it is strongly influenced by a
crossing of different liquid-glass-transition lines \cite{Sperl2004}. 

For $\mu_3=0$, an $A_4$-singularity is found; the expansion for one-component
models in Eqs.~(\ref{eq:gm}--\ref{eq:c53},\ref{eq:sol_n}) becomes invalid and 
has to be replaced by Eqs.~(\ref{eq:A4:gm}) and~(\ref{eq:A4sol}). The general 
solution in Eq,~(\ref{eq:finalA4}) has
similar properties as mentioned above for the $A_3$-singularity. Now it is the
characteristic parameter $\mu_4$ that determines how satisfactory the
approximation can be. While $\mu_4=1$ in Fig.~\ref{fig:F123} and $\mu_4=1.53$
in Fig.~\ref{fig:BKA4} allows for a description in the schematic 
models considered, the small parameter $\mu_4\approx 0.1$ 
in the microscopic models for systems with short-ranged attraction prevents the
application of the asymptotic formula.

An understanding of the critical decay law is a prerequisite for estimating
the range of validity of the Vogel-Fulcher-type laws which describe the
asymptotic limit of the time scale of the logarithmic decay laws near the 
higher-order singularities \cite{Goetze1989d}. For the two-component model 
analyzed above, the asymptotic limits were demonstrated for reasonable windows 
in time \cite{Sperl2004b}. For the mentioned colloid models, the small values
of the characteristic parameters $\mu_3$ and $\mu_4$ together with the manifest
violation of the factorization property restrict such laws to unreasonably long
times.

\ack
Our work was supported by the DFG Grant No. Go154/13-2.

\section*{References}

\end{document}